\documentclass[conference]{IEEEtran}
\IEEEoverridecommandlockouts

\usepackage{graphicx}
\usepackage{subcaption}
\usepackage{tabularx}
\usepackage{ragged2e}
\usepackage{amsmath}
\usepackage[utf8]{inputenc}
\usepackage[english]{babel}
\usepackage{hyperref}
\usepackage{xurl}
\usepackage{wasysym}
\usepackage{multirow}
\usepackage{adjustbox}
\usepackage{array}
\usepackage{booktabs}
\usepackage{xurl}
\usepackage{xspace}
\usepackage{pdflscape}
\usepackage{arydshln}
\usepackage{mathtools}
\usepackage{wrapfig}
\usepackage{float}
\usepackage{csquotes}

\usepackage{algorithm}
\usepackage{algorithmic}

\usepackage{enumerate}
\usepackage[shortlabels,inline]{enumitem}

\newcolumntype{L}{>{\RaggedRight\arraybackslash}X}
\newcommand{\coin}[1]{\texttt{#1}}
\newcommand{\tokenaddress}[2]{\href{https://etherscan.io/token/#2}{\texttt{#1}}}

\addto\extrasenglish{%
}	
\addto\extrasenglish{%
}
\addto\extrasenglish{%
}
\addto\extrasenglish{%
}

\begin{document}

\title{SoK: Yield Aggregators in DeFi}

\author{\IEEEauthorblockN{1\textsuperscript{st} Simon Cousaert}
\IEEEauthorblockA{\textit{Centre for Blockchain Technologies} \\
\textit{University College London}\\
London, WC1E 6BT, UK\\
simon.cousaert@outlook.com}
\and
\IEEEauthorblockN{2\textsuperscript{nd} Jiahua Xu}
\IEEEauthorblockA{\textit{Centre for Blockchain Technologies} \\
\textit{University College London}\\
London, WC1E 6BT, UK\\
jiahua.xu@ucl.ac.uk}
\and
\IEEEauthorblockN{3\textsuperscript{rd} Toshiko Matsui}
\IEEEauthorblockA{\textit{Department of Computing} \\
\textit{Imperial College London}\\
London, SW7 2AZ, UK\\
t.matsui19@imperial.ac.uk}
}

\IEEEoverridecommandlockouts
\IEEEpubid{\makebox[\columnwidth]{978-1-6654-9538-7/22/\$31.00~\copyright2022 IEEE \hfill} \hspace{\columnsep}\makebox[\columnwidth]{ }}

\maketitle

\begin{abstract}

Yield farming has been an immensely popular activity for cryptocurrency holders since the explosion of Decentralized Finance (DeFi) in the summer of 2020. In this Systematization of Knowledge (SoK), we study a general framework for yield farming strategies with empirical analysis. First, we summarize the fundamentals of yield farming by focusing on the protocols and tokens used by aggregators. We then examine the sources of yield and translate those into three example yield farming strategies, followed by the simulations of yield farming performance, based on these strategies. We further compare four major yield aggregators---Idle, Pickle, Harvest and Yearn---in the ecosystem, along with brief introductions of others. We systematize their strategies and revenue models, and conduct an empirical analysis with on-chain data from example vaults, to find a plausible connection between data anomalies and historical events. Finally, we discuss the benefits and risks of yield aggregators.

\end{abstract}

\begin{IEEEkeywords}
DeFi, decentralized Finance, yield aggregator, yield farming, protocols, pools, strategies, simulation
\end{IEEEkeywords}

\section{Introduction}
\label{sec:intro}

\subsection{Background}
\label{subsec:backgr}

Decentralized Finance (DeFi) has gained significant traction during the past year. In March 2020, the total value locked (TVL) in DeFi protocols was around 600 million USD \cite{defillama2021}. The total value locked indicates the amount of funds that are held by smart contracts related to DeFi protocols. That number exploded in the year coming, kick-started by the so-called ``DeFi Summer'' where TVL reached around 10 billion USD by the end of September 2020. As of November 2021, TVL numbers have reached more than 100 billion USD on Ethereum alone. Apart from the non-custodial aspect, the permissionless nature and the openly auditable protocols, a main driver for the growth of this industry is the composability of financial services. The so-called ``DeFi Stack'' or ``Money Legos'' allow protocols to build and combine functionalities, without the need of having this expertise in-house.
This means that developers can
focus on the core business, knowing that key infrastructure and money legos like decentralized exchanges, lending markets and yield services, are readily available to plug in. The result has been an explosion of innovation.

One of the applications that has come from the above-described movement is ``yield farming'', where investors passively earn yield
by transferring tokens to a wide range of yield generating smart contracts. The concept, first introduced by Synthetix \cite{snx2019liqmining}, started to gain traction on 15 June 2020 \cite{Leshner2020CompoundGovernance} with the introduction of \coin{COMP}. \coin{COMP} is the governance token (see \ref{subsubsec:nagive_gov-tokens}) of Compound, one of the major DeFi lending protocols. Compound participants get rewarded with newly-minted \coin{COMP} tokens through both lending and borrowing activities, a process termed ``liquidity mining''.
Thus, on top of the inherently designed benefit that users get for providing liquidity in different kinds of pools (e.g. interest in the case of lending protocols, or fees in the case of providing liquidity in Automated Market Making (AMM) pools), additional governance tokens are rewarded to users to further encourage their participation in the issuing platform during the early stage of adoption.
The basic yield farming idea was born: the search for opportunities in the DeFi ecosystem to generate returns on otherwise idly-sitting crypto assets. As a reaction to the creation of a multitude of platforms returning interests, fees and token rewards, yield aggregators built on top of the DeFi primitives came into existence.

\subsection{Contribution}
\label{subsec:contr}

This paper presents a systemic analysis of the parts and mechanics behind yield aggregation strategies. We identify the risks of yield farming strategies and the sources of yield. Based on an inspection of the implementation of four major yield aggregator protocols---Idle Finance \cite{idle2021impl}, Pickle Finance \cite{pickle2021impl}, Harvest Finance \cite{harvest2021impl} and Yearn Finance \cite{yearn2021impl}---we synthesize a model of the typical workflow in a yield farming strategy. Based on this workflow and three frequently adopted strategies, we simulate the wealth of a yield aggregator in a controlled market environment. Finally, major yield farming protocols are compared against each other on both theoretical and empirical grounds.
\section{Fundamentals of yield farming}
\label{sec:taxonomy}

A yield aggregator is a set of smart contracts (called a ``protocol'') that pools investors' funds, and invests them in an array of yield-producing products or services through interacting with their respective protocols (see \ref{subsec:protocols}). The process of generating yield on crypto assets is termed ``yield farming'', and yield aggregator users are thus also known as ``yield farmers''.
In essence, a yield aggregator can be deemed as a smart contract-based fund manager, whose investment strategies are pre-programmed and automatically executed~\cite{Xu2022b}.

\subsection{Related protocols}
\label{subsec:protocols}

Yield farming relies on other building blocks in the DeFi stack. In this section, we outline the essential types of protocols that are used within yield farming strategies and shortly explain the mechanics behind them. 

\subsubsection{Lending platforms}
\label{sec:plf}

Lending platforms enable lending and borrowing of on-chain assets,
with the interest rate set programmatically by a smart contract \cite{Gudgeon2020PLF}.  On these platforms, funds are pooled together to be borrowed by other users that want to take out a loan in a specific asset.
Suppliers of loanable funds receive interest over time, borrowers pay interest. Generally speaking, interest in a given market is accrued through market-specific, interest-bearing derivative tokens \cite{Perez2020liquidations}. The most prominent lending platforms are Aave \cite{Aave2021}, Compound \cite{Compound2021} and dYdX \cite{dydx2021}. 

\subsubsection{Automated Market Makers (AMM)}
\label{sec:amm}

AMMs provide liquidity algorithmically by pooling funds and determining asset prices via a so-called conservation function \cite{xu2021dexAmm}.
In a liquidity pool, reserves of two or more assets are locked into a smart contract. There are two core actors involved in the general mechanism of AMMs. First, there are the liquidity providers, who add funds to the asset pool. In return, they receive liquidity provider (LP) tokens proportionate to their liquidity contribution as a fraction of the entire pool. Generally, these LP tokens can be redeemed anytime the liquidity provider wants to get his funds back. The second core actors are the traders, who swap an input asset (what to be spent) for an output asset (what to be received). The exchange rate between the assets is deterministic, depending on the size of the pool and the desired amount to be traded. 
The most prominent AMMs are Uniswap \cite{Uniswap2021}, Balancer \cite{Balancer2021}, Curve \cite{Curve2021} and Bancor \cite{Bancor2021}.

\subsubsection{Other}
Next to lending platforms and AMMs, which are the backbone of DeFi, other protocols can offer yield in their own way. This can be through the products they offer, or through liquidity mining programs. Examples of yield offering products that are neither lending platforms, nor AMMs, are Synthetix \cite{synthetix2021home}, Lido \cite{lido2021home} and Pooltogether \cite{pooltogether2021home}.




\subsection{Tokens}
\label{subsec:tokens}

While the protocols in \autoref{subsec:protocols} are smart contracts that pragmatically define the mechanisms behind the concept, tokens are smart contracts that generally follow a standard token interface, representing assets, synthetic assets or derivatives \cite{Werner2021sokDefi}. In this section, we discuss the most important tokens in yield farming.

\subsubsection{Stablecoins}
\label{subsubsec:stablecoins}

Stablecoins are cryptocurrencies that attempt to offer price stability. They are designed to have a stable price with respect to some reference point \cite{moin2019stablecoin}, being another cryptocurrency, fiat money, or even commodities. Today's biggest stablecoins, {\tt USDT}, {\tt USDC} and {\tt BUSD} are all pegged to the US Dollar. Stablecoins can be non-custodial or custodial, asset-backed or algorithmically programmed. For a more thorough discussion on stablecoin design, we direct the reader to \cite{klages2020stablecoins}. The characteristic of potential lower volatility compared to traditional cryptocurrencies, makes stablecoins a desirable class of assets for yield farming. 

\subsubsection{Lending platform interest-bearing tokens}

As discussed in \autoref{subsec:protocols}, when lenders supply funds in a market pool, they receive interest-bearing tokens that represent the lender's share in the pool. Over time, borrowers pay interest that flows into the pool, increasing the value of the lender's relative share. This mechanism ensures that accrued interest in a market is paid out to the fund providers. Examples of lending platform interest-bearing tokens are aTokens by Aave and cTokens by Compound.

\subsubsection{AMM liquidity provider (LP) tokens}

Whenever liquidity is deposited into a pool, liquidity tokens are minted and sent to the liquidity provider. Similar to interest-bearing tokens in lending platforms, AMM liquidity tokens represent a proportional share of the pool and the same mint and burn mechanisms are applied to keep track of liquidity. One key difference is that there are no borrowers, but traders, who pay a fee each time they do a trade. For example, Uniswap charges a 0.3\% fee. Every time a user swaps an asset in a certain pool, 0.3\% of the purchased amount is retained to the pool and thus distributed pro-rata to all liquidity providers. Sushiswap charges a slightly lower fee of 0.25\%.






\subsubsection{Native/Governance tokens}
\label{subsubsec:nagive_gov-tokens}

Liquidity mining is a big part of yield farming strategies. DeFi projects are in need of liquidity as more funds often equal higher revenue streams. Both early and more mature projects therefore launch liquidity mining programs, incentivizing users to supply liquidity into the protocol (see \ref{subsec:native_tokens}). These free bonus rewards are typically an attractive part of a yield-generating strategy and it encourages yield farmers to look for the most attractive strategies in both the short and long term \cite{berenzon2020liqmining}. Examples of governance tokens as part of a liquidity mining program are \coin{COMP} from Compound, \coin{UNI} from Uniswap, \coin{SUSHI} from Sushiswap and \coin{YFI} from Yearn Finance. 


\subsection{Yield}
\label{sec:yield_origin}

\begin{figure}[ht]
\centering
\includegraphics[width=55mm,scale=0.5]{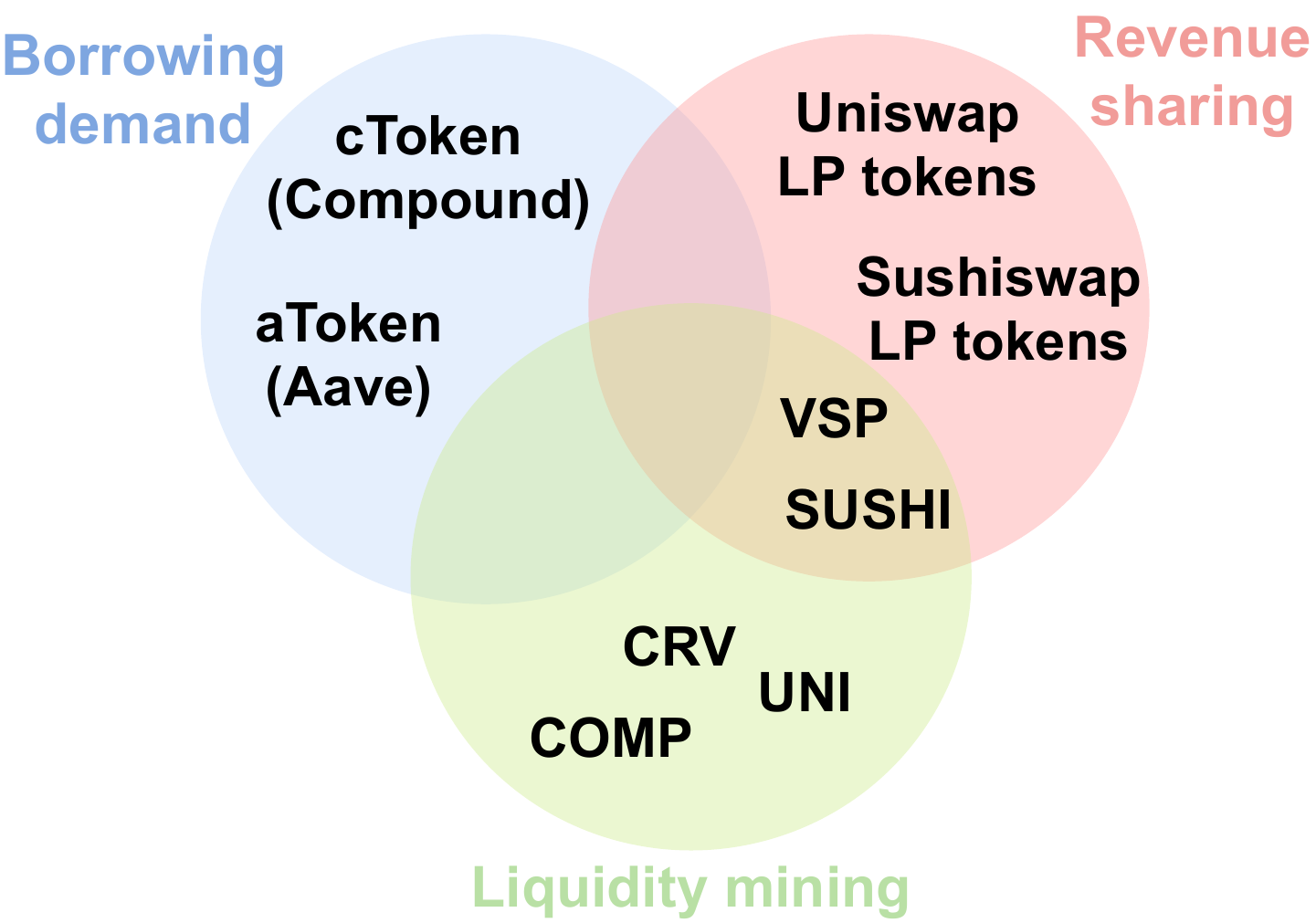}
\caption{Yield sources and representative crypto assets associated with each source. \coin{cTokens} and \coin{aTokens} are interest-bearing tokens from the lending markets Compound and Aave respectively. Uniswap and Sushiswap LP tokens are tied to their respective AMMs. \coin{COMP}, \coin{UNI} and \coin{CRV} are governance tokens that have been or are currently being distributed to protocol users of Compound, Uniswap and Curve, respectively, as part of the liquidity mining program. \coin{VSP} and \coin{SUSHI} are governance tokens of Vesper and Sushiswap distributed in a liquidity mining program, as well as tokens entitled to revenue sharing when they are staked within their protocol.}
\label{fig:yield_assets}
\end{figure}

Yield, defined as ``the total amount of profit or income produced from a business or investment'' \cite{Dictionary2021yield}, is often measured in terms of Annual Percentage Yield (APY). Yield is the ultimate goal pursued by farmers, and typically originates from borrowing demand, liquidity mining, as well as revenue sharing (\autoref{fig:yield_assets}).




\subsubsection{Borrowing demand}
\label{subsec:borrowing_demand}

As the demand for loans in crypto assets grows, the borrowing interest rate increases, leading to higher yields for lenders.
Particularly in a bullish market, speculators are keen to borrow funds despite a high interest rate, in expectation of an appreciation in the assets of their leveraged long position.
A borrower wishing to increase their exposure to \coin{ETH}, for example, may use \coin{ETH} as collateral to borrow \coin{USDC}, then repetitively exchanging \coin{USDC} for \coin{ETH} to deposit it as collateral to borrow more \coin{USDC}, forming a ``leveraging spiral'' \cite{Xu2021c}.
Compound and Aave, two major DeFi lending protocols (see \autoref{sec:plf}), have witnessed the borrow APY of \coin{USDC} rising from 2-3\% in May 2020 to as high as 10\% in April 2021.\footnote{\url{https://app.defiscore.io/assets/usdc}} 
This specific kind of yield is incorporated in interest-bearing tokens, such as \coin{cTokens} from Compound or \coin{aTokens} from Aave. 

\subsubsection{Liquidity mining}
\label{subsec:native_tokens}

Another yield source comes from liquidity mining programs, where early participants receive native tokens representing protocol ownership. This incentivizes people to contribute funds into the protocol, and enhances decentralization as the protocol ownership is distributed to users. 
The native tokens often have a governance functionality attached to them which is deemed valuable, as the token holders have a say in the future strategic direction of the project. Native tokens sometimes also entitle holders to a share of the protocol revenue (see the below). Further, the values these tokens possess itself especially in a speculation context can be the benefits of owing a protocol.



\subsubsection{Revenue sharing}
\label{subsec:revenue_sharing}

Some tokens entitle users to part of the revenue that is going through the protocol. These can be governance tokens or other kinds of tokens. 
One example is the liquidity provider tokens in AMM-based DEXs \cite{xu2021dexAmm}. By supplying liquidity into an AMM pool, users receive the fees that are paid by traders within that pool. The higher the volume in that pool, the more fees that are generated, and the more a liquidity provider profits from this. In Uniswap \cite{Uniswap2021}, a 0.3\% fee is charged for every trade within a pool and goes fully to LPs. 

This brings up a second kind of revenue-sharing token, where users have to actively stake their tokens to receive a share of the revenue. For example, \coin{SUSHI} holders that stake their \coin{SUSHI} will get \coin{xSUSHI} in return, which represents the proportional share of a pool that captures 0.05\% of all trades on Sushiswap \cite{xsushi2021faq}. Vesper Finance's governance token, \coin{VSP}, can also be deposited in a pool, in return for \coin{vVSP}, a token that represents the user's proportional share of a pool that captures part of the revenue generated throughout the whole Vesper platform~\cite{vesper2021tokenomics}.

\section{Yield Farming Strategies}
\label{sec:formalization}

A yield farming strategy is made of a specific set of actions through modular smart contracts that automates the yield farming process. One pool (also called ``vault'' or other names) commonly employs only a single strategy, but more recent developments allow for multiple strategies within one asset pool, such as the Yearn v2 vaults discussed in \ref{subsec:yearn}.

\subsection{Pool workflow}

\begin{figure}[h!]
\centering
\includegraphics[width=\linewidth]{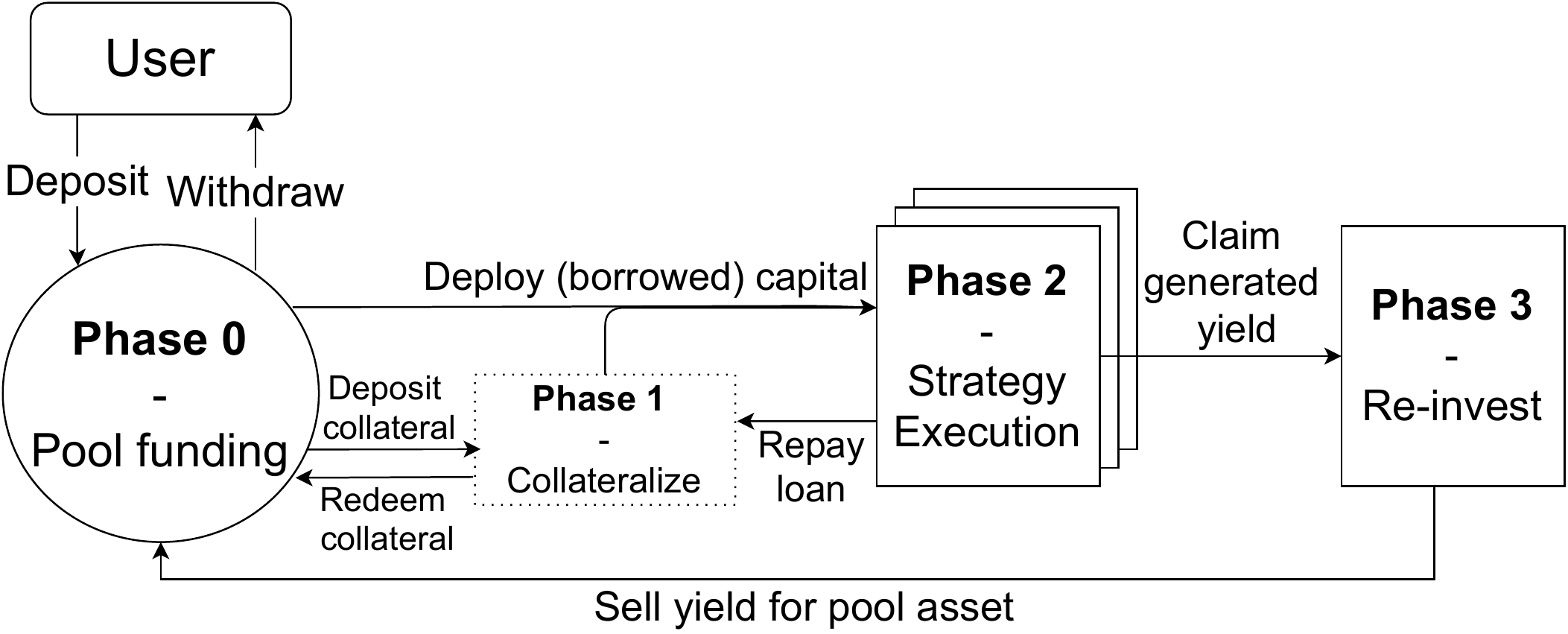}
\caption{Stylized yield aggregator mechanism}
\label{fig:mechanism_all}
\end{figure}

A typical yield aggregator pool workflow has three to four phases, depending on the specific protocol. \autoref{fig:mechanism_all} visualizes the typical workflow within yield aggregator pools. Note that the steps in this figure are not necessarily sequential, but can happen in random order once the pool has been bootstrapped.

\begin{figure}
\centering
\includegraphics[width = \linewidth]{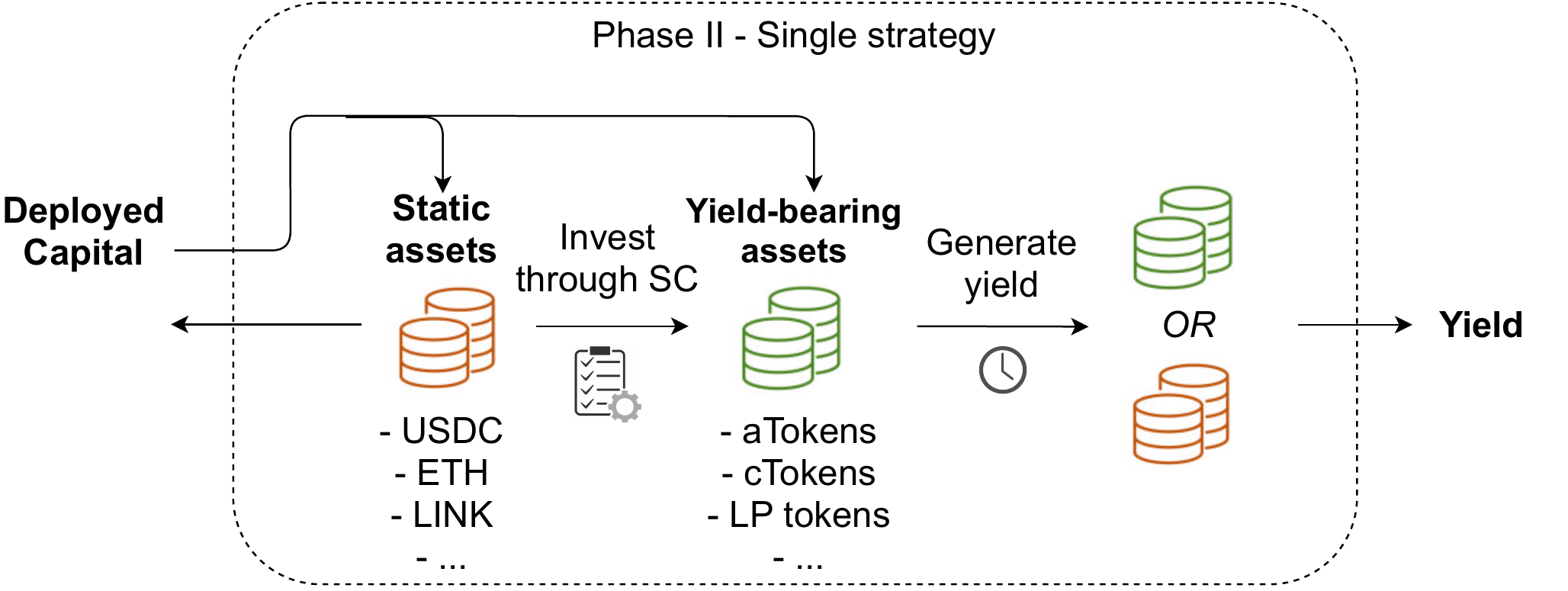}
\caption{Execution process of a single strategy. SC = Smart Contract}
\label{fig:mechanism_phase2}
\end{figure}

\begin{enumerate}[start=0,label={\bfseries Phase~\arabic*}, wide]
\item
\label{item:phase0}
In this phase, a yield farming strategy is proposed, created and deployed on the blockchain. 
Upon approval of the strategy---either by the protocol development team, or through a decentralized governed voting mechanism, a pool is deployed in order to collect and disperse funds. 
From that moment, yield farmers can deposit funds into the pool or  withdraw from it.
By depositing funds, farmers receive pool shares in forms of ``liquidity tokens''. By withdrawing funds, they surrender those liquidity tokens to redeem funds proportionate to their shares. 

\item
\label{item:phase1}
In this phase, the funds pooled in \ref{item:phase0} are used as collateral to borrow another asset through lending platforms like Aave \cite{Aave2021} or Compound \cite{Compound2021}. 
The purpose of this phase is to turn miscellaneous assets collected from farmers into another asset, typically a stablecoin, to prepare for strategy execution in \ref{item:phase2}.
Naturally, this phase can be skipped, if the pooled asset already satisfies the asset type requirement. For example, if the strategy only accepts \coin{DAI}---a stablecoin pegged to USD---as the deployable capital, then funds in \coin{ETH} and \coin{WBTC} pools need be deposited to a lending platform (see \ref{sec:plf}) as collateral to borrow \coin{DAI} first before proceeding to \ref{item:phase2}.

A critical aspect in this stage is collateral management to avoid liquidation (see \ref{subsubsec:liquidationrisk}) of deposited assets. The yield aggregator manages the collateral risk level by directing the flow of funds between smart contracts used in \ref{item:phase0} and \ref{item:phase1}. 


\item 
\label{item:phase2}
In this phase, funds from \ref{item:phase0} and/or \ref{item:phase1} are deployed to generate yield, by following the pre-programmed strategy (for examples see \ref{subsec:examples}). 
The output of \ref{item:phase2} is the generated yield. 
\autoref{fig:mechanism_phase2} shows how a single strategy typically works. ``Static'' assets such as \coin{USDC} and \coin{WETH}, represented by red coins, constitute the value-denominating underlying of yield-bearing assets such as \coin{cUSDC} and \coin{aWETH} (see \autoref{fig:yield_assets}), which are represented by green coins. Depending on the pool, assets that are supplied can already be yield-bearing, such as LP tokens. These kinds of pools typically skip \ref{item:phase1}. Note that it is also possible for borrowed assets to flow back to \ref{item:phase1}, in order to pay back part of the loan in case of a relatively big fund withdrawal from the original pool. 
\item
\label{item:phase3}
The goal of \ref{item:phase3} is to get the generated yield back to the original fund.
Depending on which assets constitute the yield, they can be exchanged on a DEX for 
the original asset and returned to \ref{item:phase0}, or deposited in \ref{item:phase1} as collateral to borrow more deployable capital, or, when acceptable to the strategy, reinvested directly in \ref{item:phase2}.
\ref{item:phase3} can be triggered after a certain time, or after a specific smart contract function from which the previous phases is called. It is executed either automatically or manually, depending on the aggregator \cite{yearn2021vaultFunctions} \cite{harvest2021vaultFunction}. 

\end{enumerate}

\subsection{Common strategies}
\label{subsec:examples}

In this section, we describe three common yield farming strategies: {\it simple lending},
{\it leveraged borrow},
and {\it liquidity provision},
with illustrations of their mechanisms (\autoref{fig:example_strategies}). 
We additionally simulate the expected performance of each strategy in a controlled, parameterized environment, by tracking the trajectory of the total value $W$ of the yield aggregator.\footnote{The code repository can be found \href{https://github.com/SimonCousaert/yieldAggregators}{here}.}

\subsubsection{Assumptions}

On comparing the three common strategies, for simple demonstration purposes without loss of generality, we make the following assumptions:

\begin{enumerate}[label=\arabic*.]
    \item the transaction cost is neglected;
    \item the value of the yield aggregator $W_t$ is measured in \coin{DAI}; at $t = 0$, the aggregator contains 1 \coin{DAI}'s worth of funds, i.e., $W_0 = 1$;
    \label{ass:farmingpoolvalue}
    \item the aggregator supplies all funds in the pool to a yield-generating protocol---either a lending platform or an AMM, and the funds represent 1\% of the protocol's total assets held at $t = 0$;
    \label{ass:2}
    \item the yield-generating protocol---either a lending platform or an AMM---distributes 0.01 governance token per day to its users proportionately to their stake in the protocol:
        \begin{enumerate}[label=\alph*., ref=\theenumi{}\alph*]
            \item for a lending platform, half of the governance tokens are distributed to lenders proportionate to their deposits, and half to borrowers proportionate to their loans,
            \label{item:plfgov}
        \item for an AMM, the governance tokens are distributed proportionately to LPs;
        \end{enumerate}
    \item the governance token price remains constant during the simulation period;
    \item the lending platform has a fixed borrow APY of 10\% and a collateral factor of 80\%, meaning for each \coin{DAI} deposited, the depositor is allowed to take 0.8 \coin{DAI}'s worth of loan; at $t=0$, 70\% of the funds in the lending platform pool is lent out, and all other market participants' additional borrow and repay, as well as deposit and withdraw cancel each other out on an aggregate level during the simulated period; 
    \label{item:borrowapy}
    \item the AMM has a fixed exchange fee of 5\% and applies a Uniswap-like constant-product conservation function;\footnote{We refer the reader to \cite{Angeris2020c} for detailed description and analyses on AMMs with constant-product conservation function.} the fee is charged by retaining 5\% of the theoretical fee-free purchase quantity within the AMM pool.
    \label{ass:amm}
\end{enumerate}

\subsubsection{Mechanism and simulation results}

\begin{figure}[tbp]
    \centering
    \begin{subfigure}[b]{\linewidth}
        \centering
        \includegraphics[width=\linewidth]{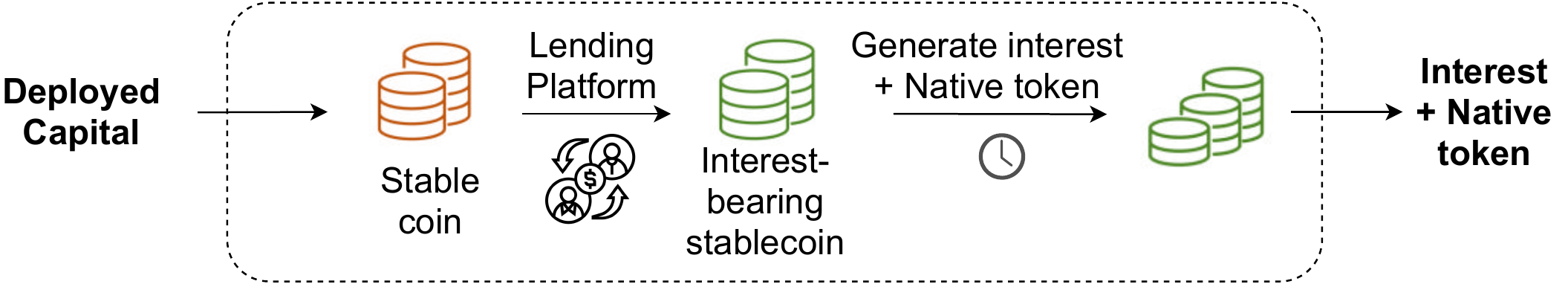}
        \vskip -0.7em
        \caption{Simple lending}
        \label{fig:simpleLending}
    \end{subfigure}
    \begin{subfigure}[b]{\linewidth}
        \centering
        \includegraphics[width=\linewidth]{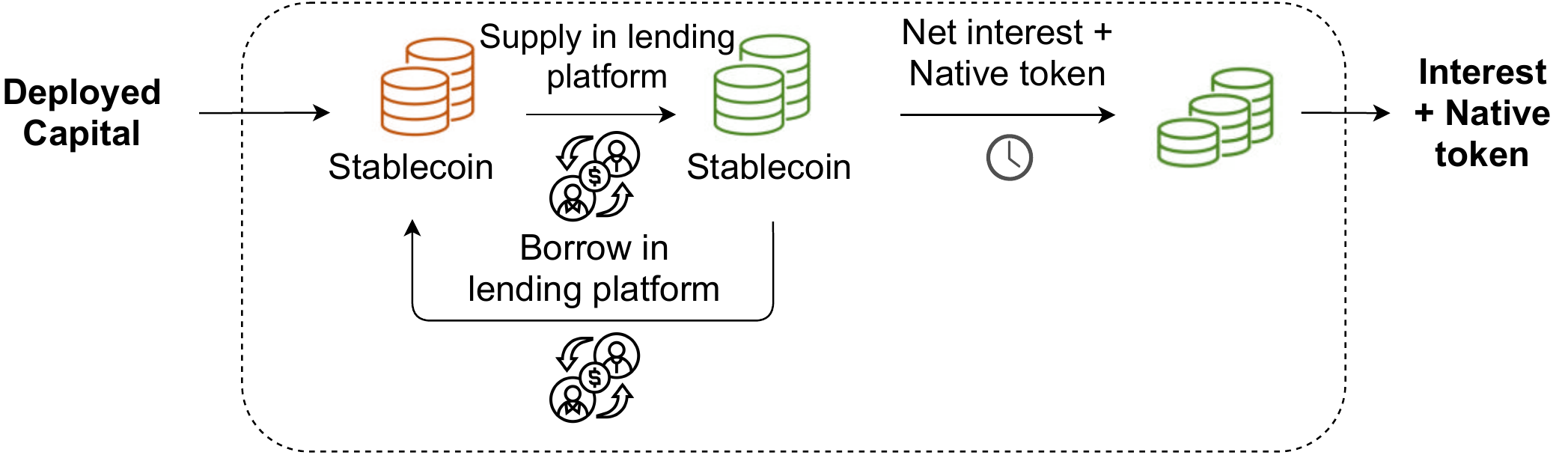}
        \vskip -0.7em
        \caption{Leveraged borrow}
        \label{fig:leveragelending}
    \end{subfigure}
    \begin{subfigure}[b]{\linewidth}
        \centering
        \includegraphics[width=\linewidth]{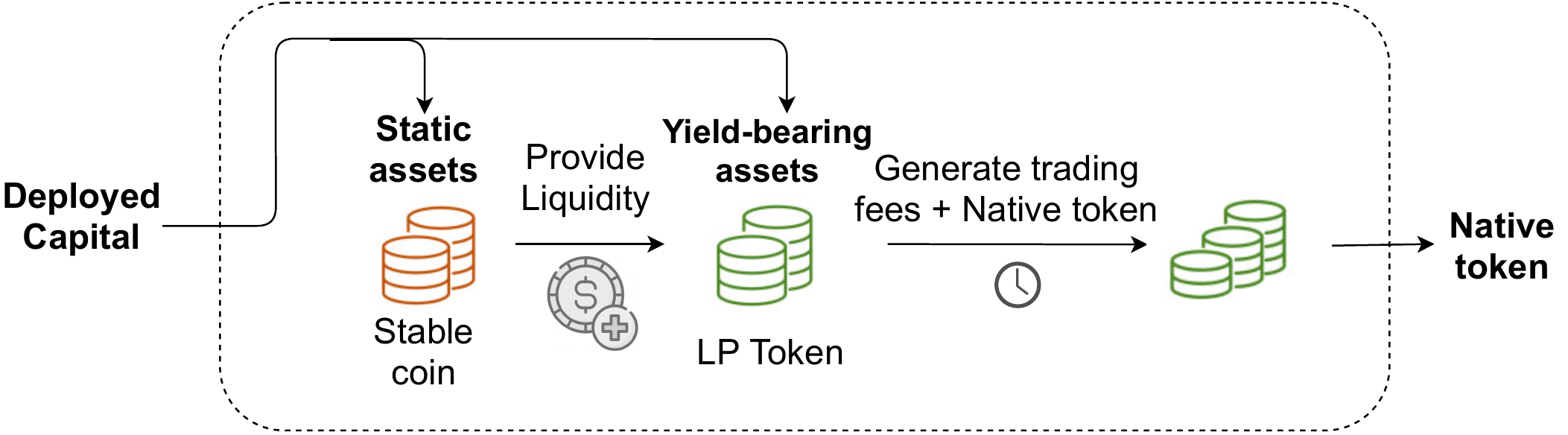}
        \vskip -0.7em
        \caption{Liquidity provision}
        \label{fig:liqmining}
    \end{subfigure}
\caption{Classic yield aggregator strategies}
\label{fig:example_strategies}
\end{figure}


\paragraph{Simple lending}
\label{sec:simplelending}

With {\bf Strategy \ref{algo:simple_lending}}, the aggregator earns yield through lending interest and reward tokens distributed by the lending platform (\autoref{fig:simpleLending}).

In our simulated environment, at $t=0$ the yield aggregator deposits 1 \coin{DAI} to a lending platform, and receives in return some \coin{interest-DAI} as a certificate of deposit. According to Assumption~\ref{ass:2}, the aggregator owns 1\% of the total circulating supply of \coin{interest-DAI}. 

The \coin{interest-DAI} holding of the aggregator is worth exactly 1 \coin{DAI} at $t=0$, and increases in value due to interest accrued with the passage of time. In addition, the aggregator is rewarded with the lending platform governance tokens owing to its contribution to the lending platform, and the value of the governance token holding is counted towards the total value of the aggregator.

\begin{figure*}[ht]
  \centering
        \begin{subfigure}{\linewidth}
        \centering
        \includegraphics[height=0.15\textheight, trim = {5, 0, 0, 0}, clip]{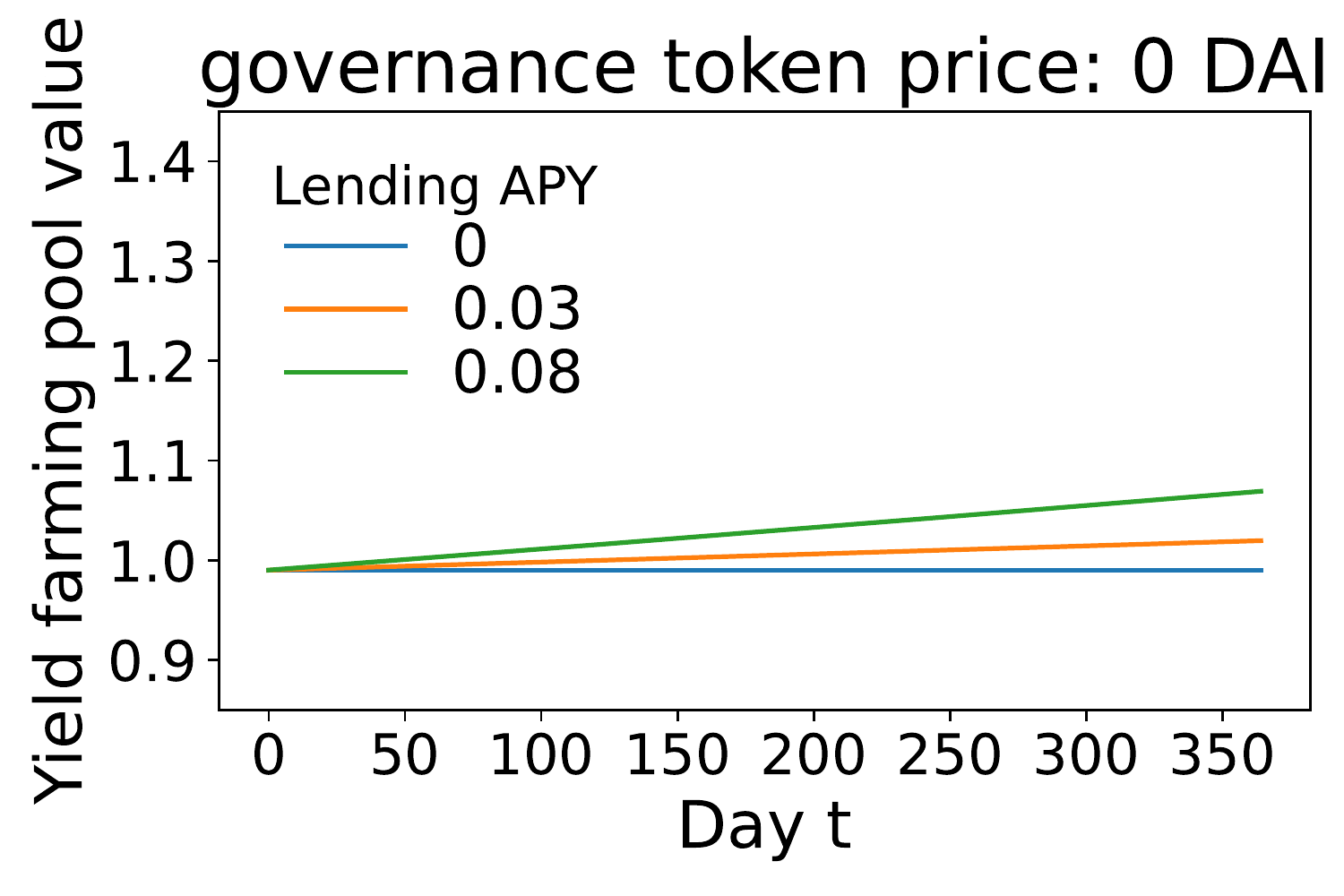}
        \includegraphics[height=0.15\textheight, trim = {65, 0, 0, 0}, clip]{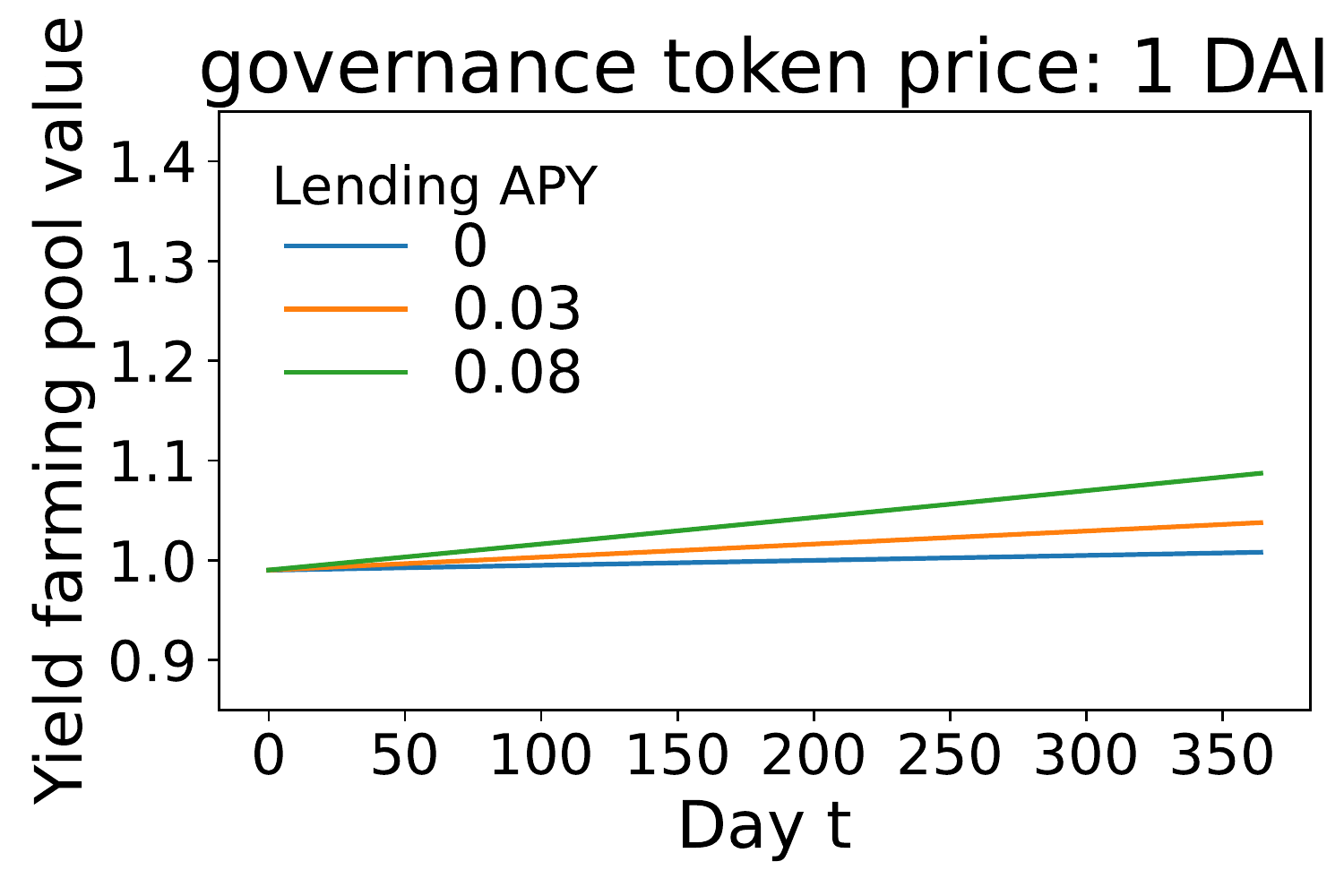}        \includegraphics[height=0.15\textheight, trim = {65, 0, 0, 0}, clip]{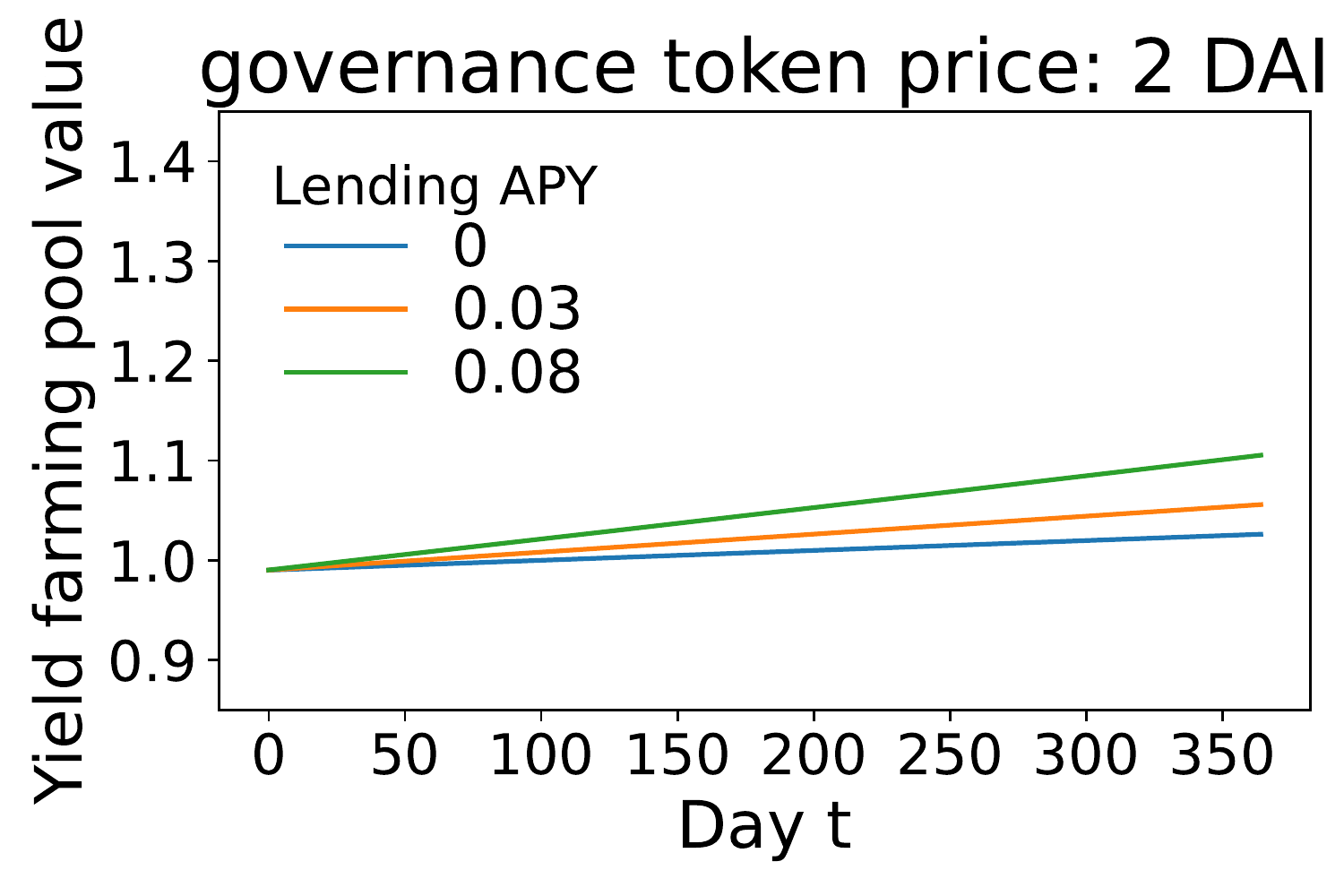}
        \vskip -0.5em
        \caption{Simple lending}
        \label{fig:simple_lending_sim}
        \end{subfigure}
        
        \begin{subfigure}{\linewidth}
        \centering
        \includegraphics[height=0.15\textheight, trim = {5, 0, 0, 0}, clip]{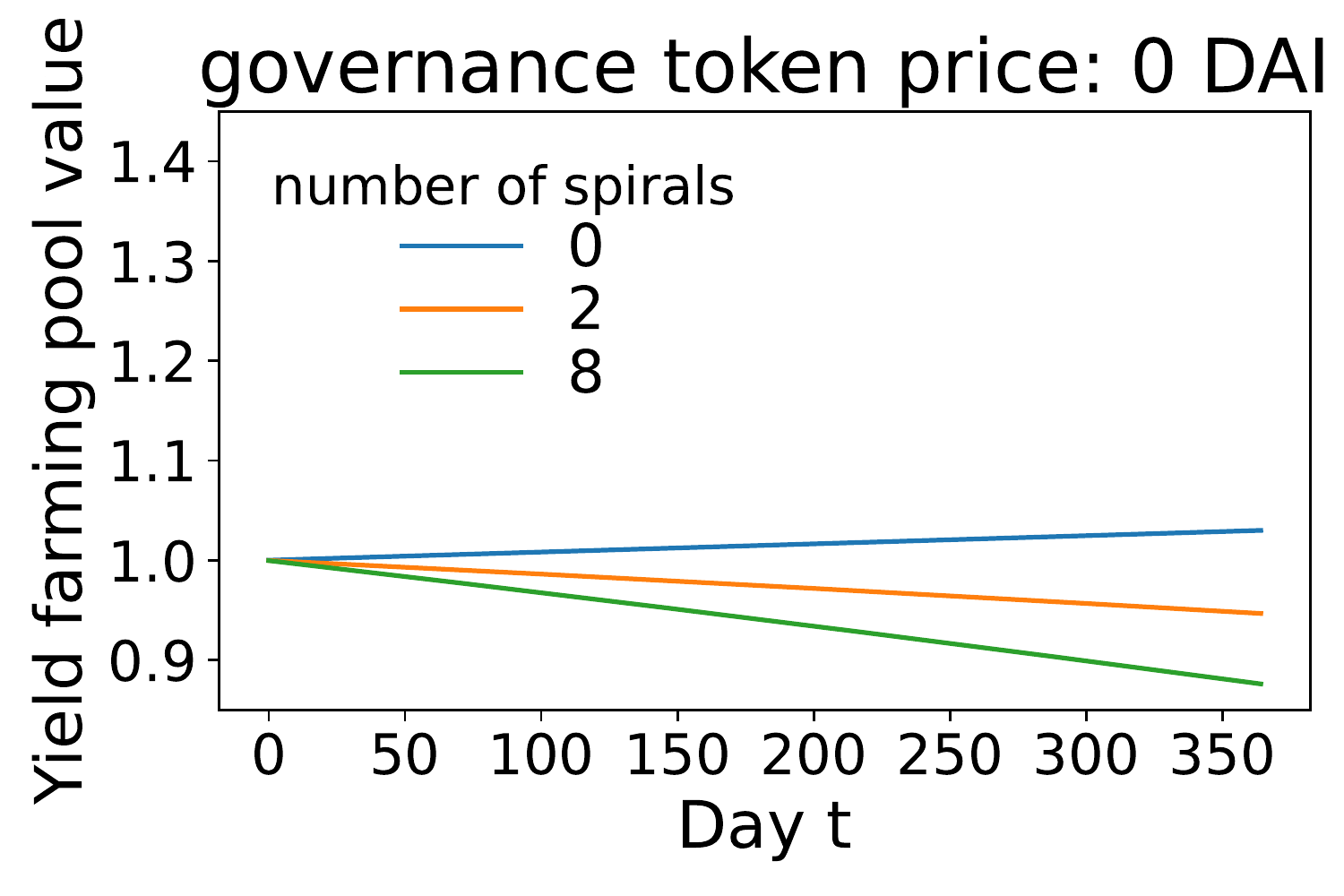}
        \includegraphics[height=0.15\textheight, trim = {65, 0, 0, 0}, clip]{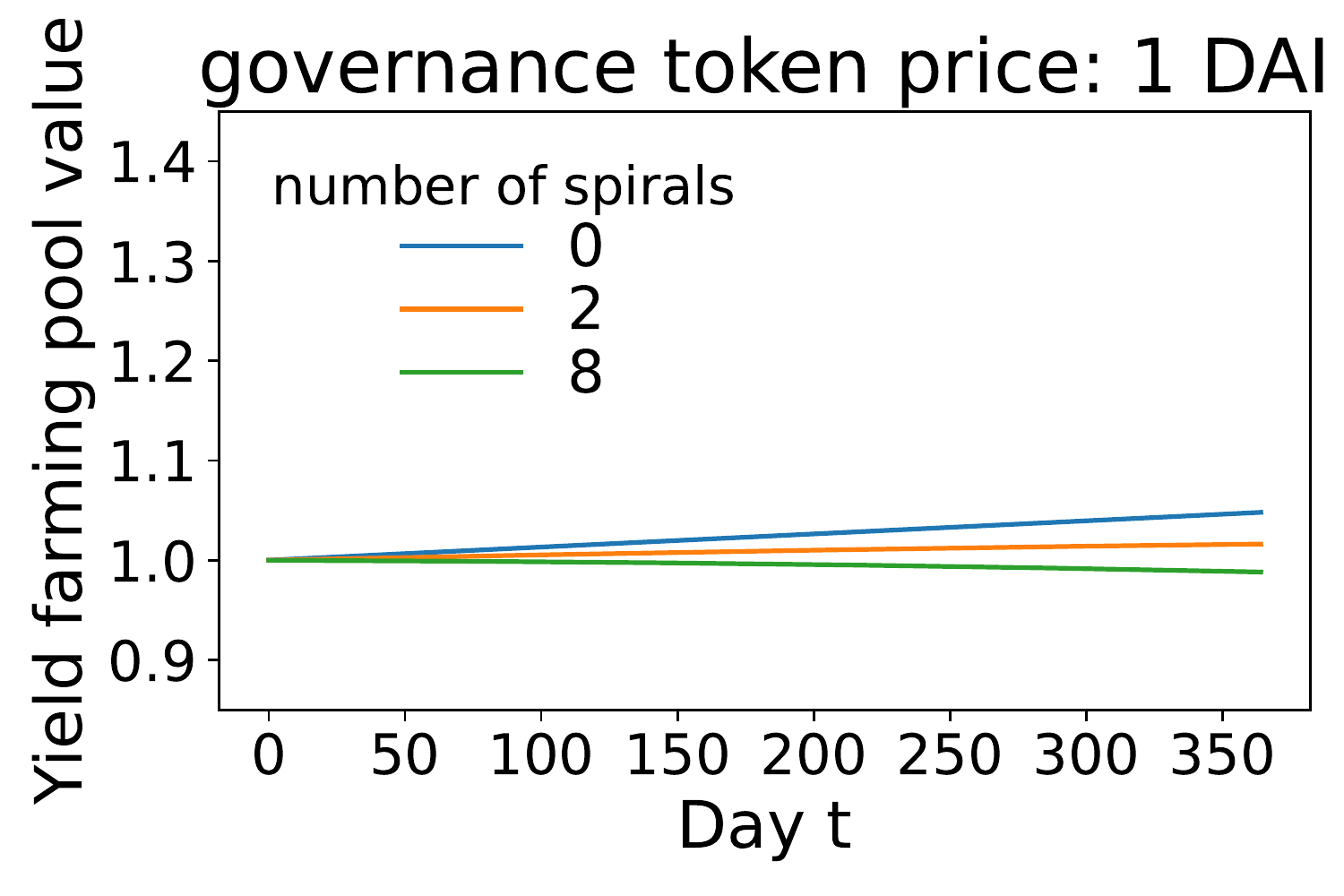}
        \includegraphics[height=0.15\textheight, trim = {65, 0, 0, 0}, clip]{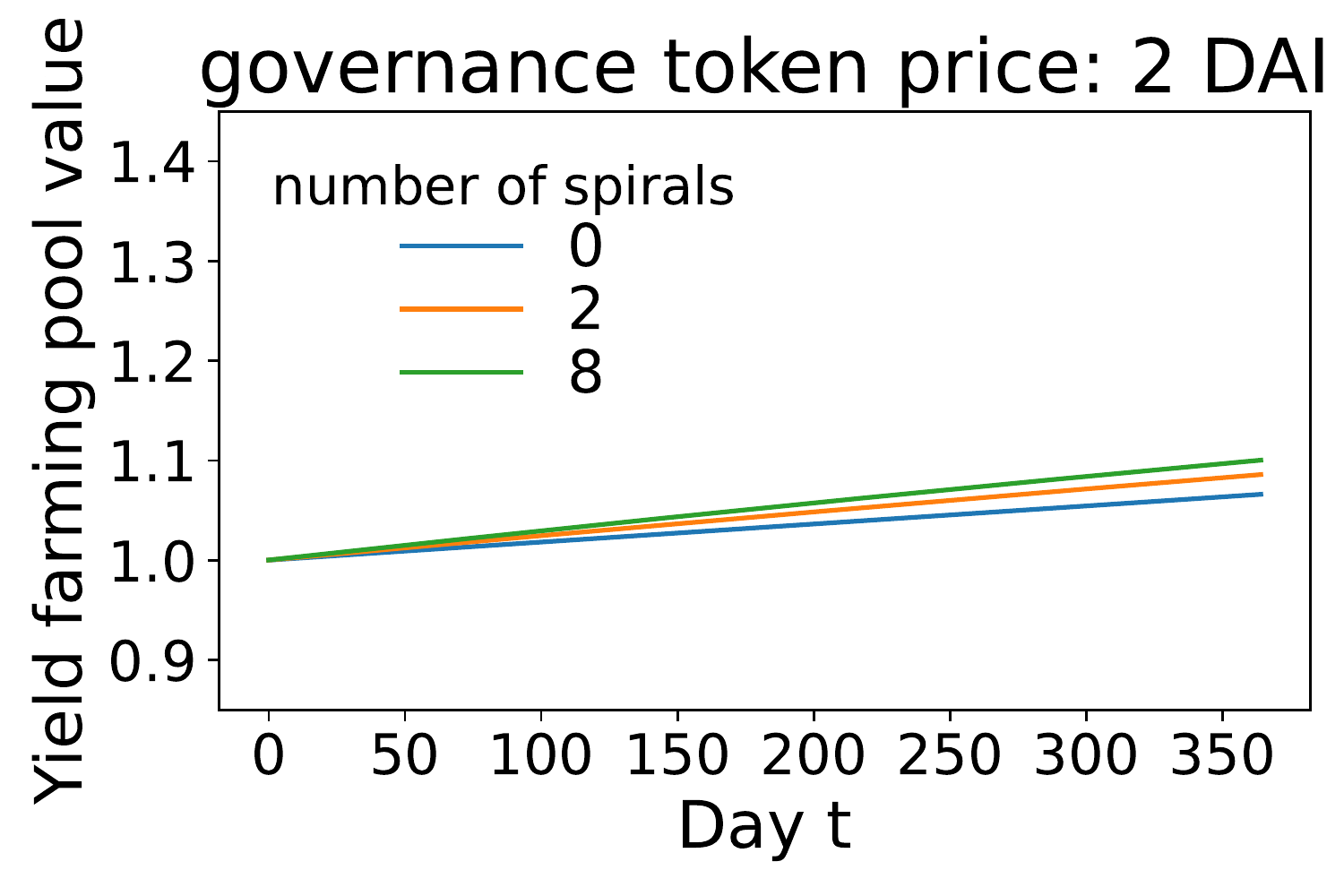}
        \vskip -0.5em
        \caption{Leveraged borrow}
        \label{fig:spiral_lending_sim}
        \end{subfigure}
        
        \begin{subfigure}{\linewidth}
        \centering
        \includegraphics[height=0.15\textheight, trim = {5, 0, 0, 0}, clip]{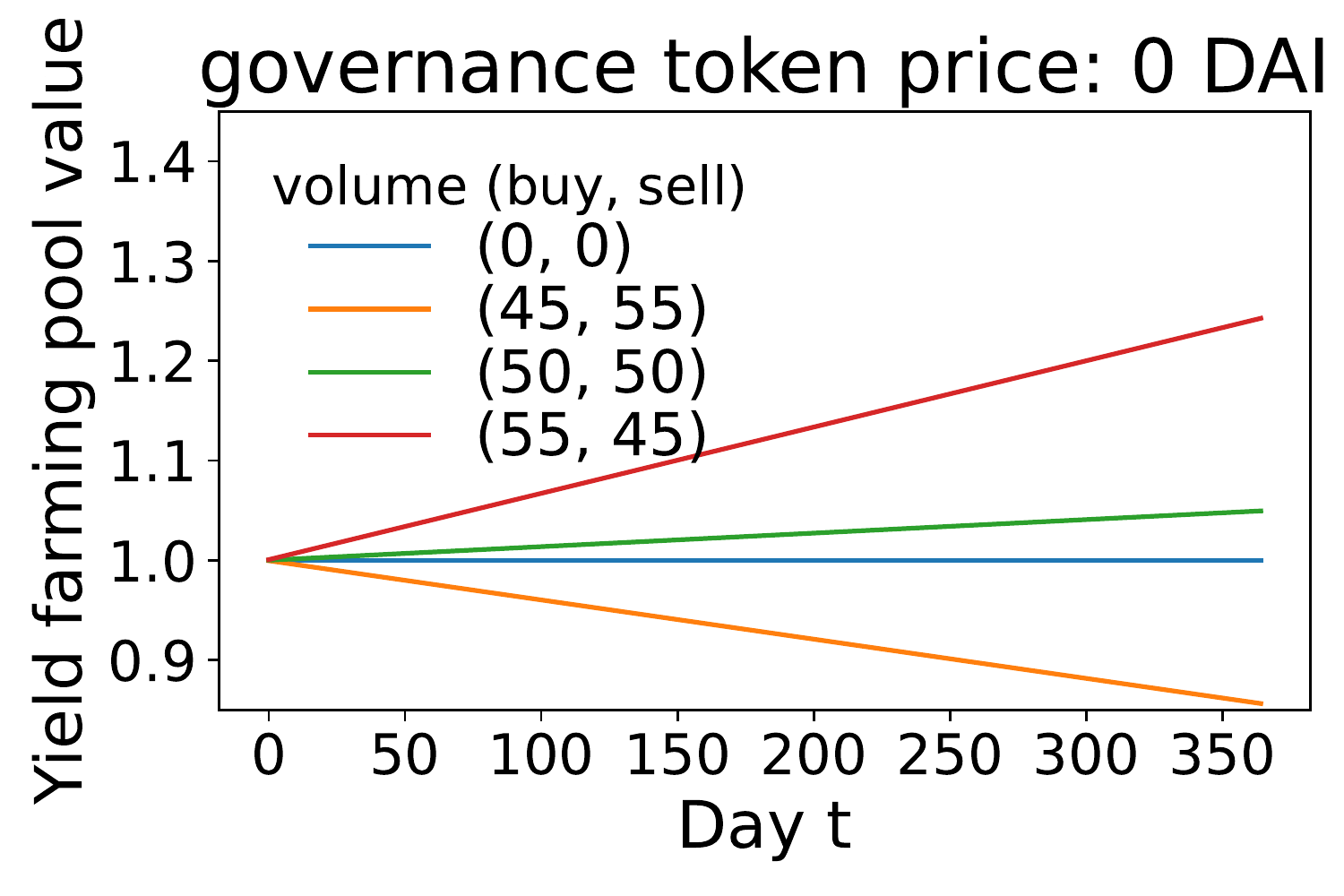}
        \includegraphics[height=0.15\textheight, trim = {65, 0, 0, 0}, clip]{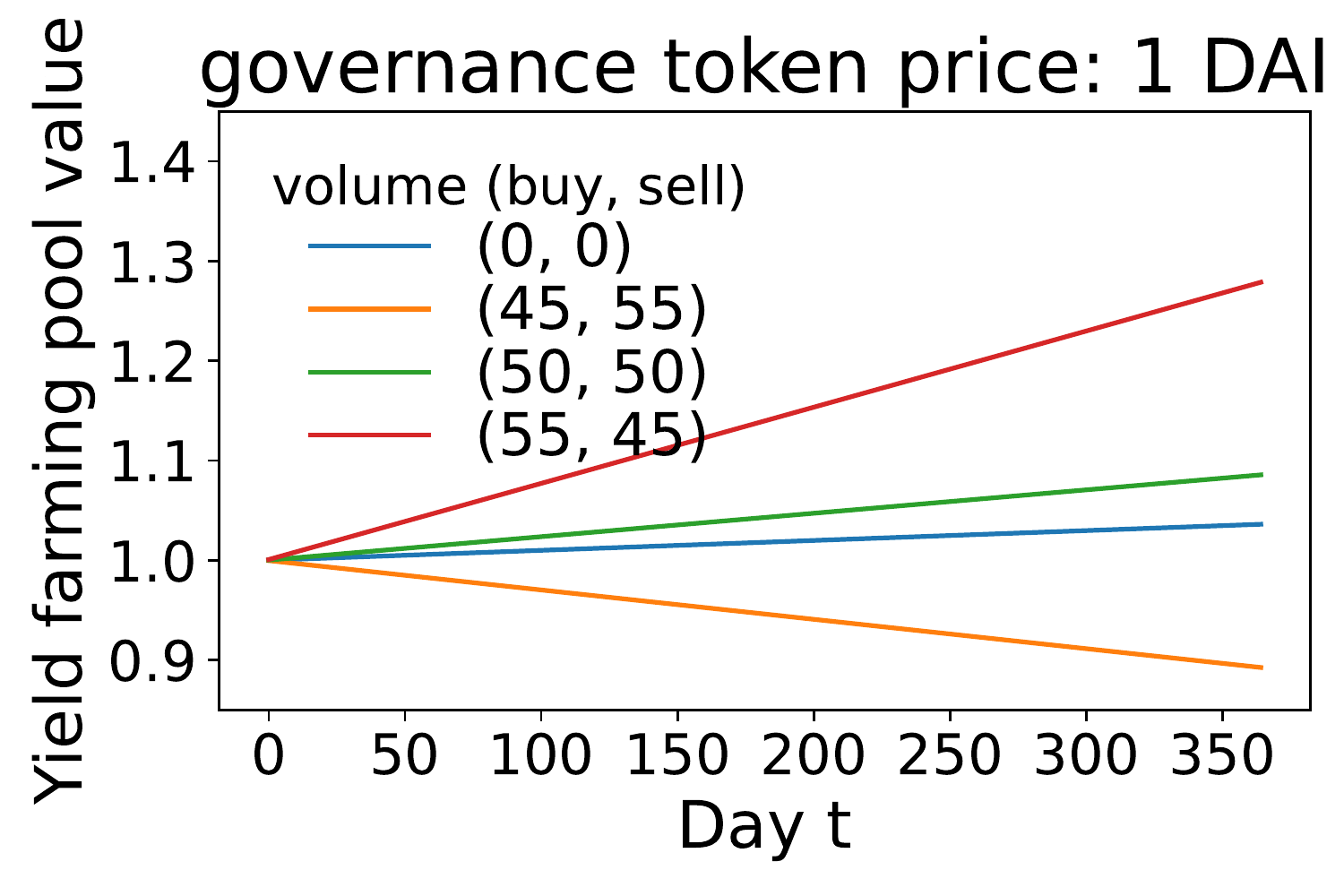}
        \includegraphics[height=0.15\textheight, trim = {65, 0, 0, 0}, clip]{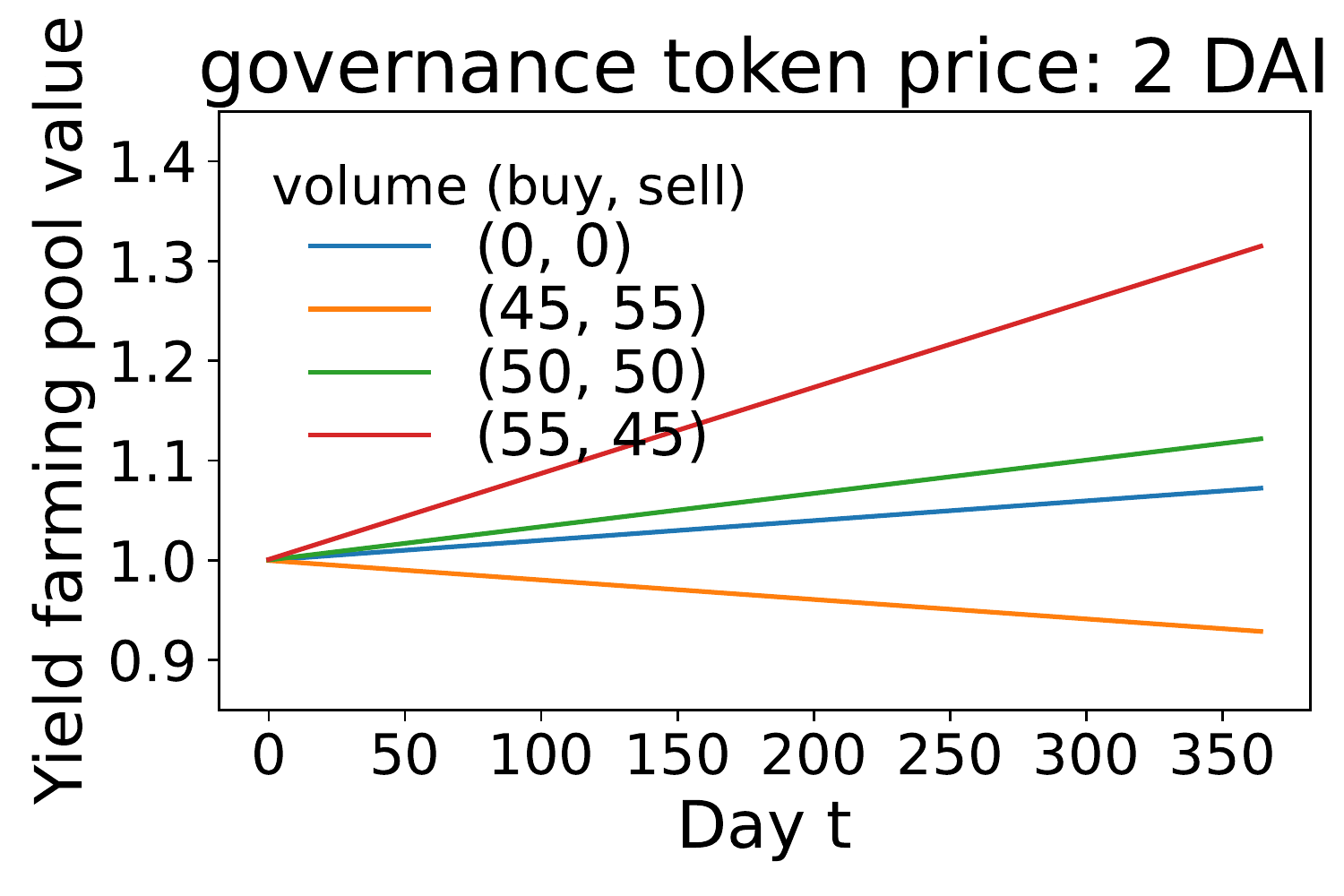}
        \vskip -0.5em
        \caption{Liquidity provision}
        \label{fig:lp_mining_sim}
        \end{subfigure}
        \caption{Simulation results}
        \label{fig:simulation_results}
\end{figure*}

The simple lending strategy is a low-risk one, and losses only occur under extreme market conditions, e.g. when the price of the asset lent out relative to the collateral suddenly increases to such a significant extent that the loan becomes undercollateralized (see lending protocol MakerDAO's Black Thursday Incident \cite{Kjaer2021,Perez2020liquidations}). Kao et al.~\cite{Kao2020a} simulate a wide range of market volatility to stress-test lending protocols such as Compound, and find that only rarely can undercollateralization occur.
\autoref{fig:simple_lending_sim} shows that, absent extreme market volatility, the value held by the aggregator $W_t$ is floored at 1 \coin{DAI}, with the worst-case scenario when the lending APY is 0 and the governance token distributed by the lending platform has 0 value. Intuitively, $W_t$ increases with lending APY and governance token price.

\begin{algorithm}[tbp]
  \small
  \floatname{algorithm}{Strategy}
  \caption{Simple lending}
  \label{algo:simple_lending}
  \begin{algorithmic}[1]
    \STATE {\bf Deposit} assets in a lending protocol.
    \STATE {\bf Accrue} supply interest and {\bf collect} native tokens over time.
    \STATE {\bf Withdraw} deposits with accrued supply interest.
  \end{algorithmic}
\end{algorithm}

\paragraph{Leveraged borrow}
\label{sec:leveragedborrow}

According to Assumption~\ref{item:plfgov} and in line with practices of major lending platforms such as Compound \cite{Compound2021}, governance tokens are rewarded to both lenders and borrowers. {\bf Strategy \ref{algo:leveraged_borrow}} thus aims to maximize the amount of governance tokens received by the lending platform through leveraging spirals (\autoref{fig:leveragelending}). 

In our simulated environment, at $t=0$ the yield aggregator first deposits 1 \coin{DAI} to a lending platform;
with this initial deposit as collateral, the aggregator then takes a loan worth 70\% of its deposit, i.e. 0.7 \coin{DAI}. 
To further augment its deposit and borrow amount for the entitlement of larger rewards, the aggregator re-deposits the borrowed funds, and use them as collateral to borrow again 70\% of the new deposit; and so on and so forth. Obviously, the more spirals the aggregator undertakes, the higher shares it holds at both the lending and the borrowing sides of the lending platform.   

As an asset's  borrow APY always exceeds its lending APY, the aggregator's loan accrues interest exponentially faster than its deposit. Assume a borrow APY of 10\% (as per Assumption~\ref{item:borrowapy}) and a supply APY of 3\%, we observe from \autoref{fig:leveragelending} that sufficiently valuable governance tokens can make the strategy profitable, but losses occur when the value of the governance tokens received is insufficient to offset the negative net interest revenue. 
Overall, a high degree of leverage, measured by the number of spirals, can amplify both the profit---in case of high-value governance tokens, as well as the loss---in case of low-value governance tokens. 

While for simplification purposes, the asset deposited as collateral and the asset borrowed are set to be the same (\coin{DAI}) in our simulation, it is worth noting that when the collateral asset and the borrowed one differ, additional liquidation risks arise (see \ref{subsubsec:liquidationrisk}).

\begin{algorithm}[tbp]
  \small
  \floatname{algorithm}{Strategy}
  \caption{Leveraged borrow}
  \label{algo:leveraged_borrow}
  \begin{algorithmic}[1]
    \STATE {\bf Deposit} assets in a lending protocol
    \STATE {\bf Borrow} assets with the deposits as collateral. \label{state:borrow}
    \STATE {\bf Deposit} borrowed assets. \label{state:redeposit}
    \STATE {\bf Repeat} steps \ref{state:borrow}--\ref{state:redeposit} multiple times.
    \STATE {\bf Accrue} interest and {\bf collect} native tokens over time.
    \STATE {\bf Swap} the native tokens into the assets borrowed.
    \STATE {\bf Repay} loans with accrued borrow interest.
    \STATE {\bf Withdraw} deposits with accrued supply interest when needed.
  \end{algorithmic}
\end{algorithm}

\paragraph{Liquidity provision}
\label{sec:liquidityprovision}
With {\bf Strategy \ref{algo:liquidity_provision}}, the aggregator supplies funds to an AMM in order to profit from both trading fees and governance tokens rewarded by the AMM (\autoref{fig:liqmining}).

In our simulated environment, at $t=0$ the aggregator 
deposits 1 \coin{DAI}'s worth of funds in the \coin{DAI-ETH} pool of an AMM, and receives in return some \coin{DAI-ETH-LP} tokens, representing its share in the AMM pool. According to Assumption~\ref{ass:2}, the aggregator owns 1\% of the total circulating supply of \coin{DAI-ETH-LP}. Given that \coin{DAI} is the denominating asset (Assumption~\ref{ass:farmingpoolvalue}), and that the \coin{DAI-ETH} pool applies a constant-product conservation function (Assumption~\ref{ass:amm}), the \coin{DAI-ETH} pool always contains \coin{DAI} and \coin{ETH} with equivalent value, and the total pool value thus equals twice the \coin{DAI} quantity in the pool.\footnote{We refer the reader to \cite{xu2021dexAmm} for a formal derivation on the pool value of a constant-product AMM.}

We additionally assume that, on an aggregate level, further liquidity provision and withdrawal cancel each other out. Hence, the aggregator's ownership of the AMM pool is neither diluted nor concentrated; that is, the value of the aggregator's \coin{DAI-ETH-LP} holding remains 1\% of the \coin{DAI-ETH} pool's value. Naturally, all other things equal, the value held by the aggregator increases with the value of the AMM governance token.

We test scenarios with different market movements. Specifically, we illustrate in \autoref{fig:lp_mining_sim} when during the entire simulation period
\begin{enumerate*}[label={(\roman*)}]
\item there is 0 trading volume (blue line), 
\label{item:0trading}
\item the buy and sell volume of \coin{ETH} is respectively 45 \coin{DAI} and 55 \coin{DAI} (orange line), 
\label{item:moreselleth}
\item the buy and sell volume of \coin{ETH} is each 50 \coin{DAI} (green line),
\label{item:5050trading}
\item the buy and sell volume of \coin{ETH} is respectively 55 \coin{DAI} and 45 \coin{DAI} (orange line). 
\label{item:morebuyeth}
\end{enumerate*}
We assume that the trading volume is evenly spread out throughout the simulation period.

\begin{algorithm}[bp]
  \small
  \floatname{algorithm}{Strategy}
  \caption{Liquidity provision}
  \label{algo:liquidity_provision}
  \begin{algorithmic}[1]
    \STATE {\bf Provide} assets as liquidity in an AMM pool.
    \STATE {\bf Collect} native tokens over time.
    \STATE {\bf Withdraw} liquidity.
  \end{algorithmic}
\end{algorithm}

Absent any trading activity---as in Scenario \ref{item:0trading}, the aggregator's yield solely comes from governance token reward.
The yield difference between Scenarios \ref{item:0trading} and \ref{item:5050trading} lies in the trading fee. By comparing the blue line and the green line in \autoref{fig:lp_mining_sim}, we clearly see that \ref{item:5050trading} results in higher yield with the presence of 5\% trading fee.

Scenario \ref{item:moreselleth} describes a market situation with higher selling pressure and consequently falling \coin{ETH} prices. The leads to an increase in the quantity of the depreciated \coin{ETH} and a decrease in the quantity of the denominating asset \coin{DAI} in the AMM pool, diminishing the AMM pool's value. When the trading fee revenue and governance token reward are insufficient to offset this value loss, the yield would be negative. 

In contrast to Scenario \ref{item:moreselleth}, Scenario \ref{item:morebuyeth} describes an opposite market situation where a higher demand in \coin{ETH} drives up its price. The leads to a decrease in the quantity of the appreciated \coin{ETH} and an increase in the quantity of the denominating asset \coin{DAI} in the AMM pool, raising the AMM pool's value.

Note that in both Scenarios \ref{item:moreselleth} and \ref{item:morebuyeth}, liquidity providers suffer from divergence loss \cite{xu2021dexAmm}, that is, they could have been better off by just holding their \coin{DAI} and \coin{ETH}, as opposed to supplying them to the AMM pool. 

The simulation shows that the liquidity provision strategy also entails risk, associated with market movements of the assets within the AMM pool. Higher volatility of the AMM pool assets implies higher uncertainty in yield.

\subsubsection{Summary} There is no free lunch---the same applies to yield farming. In this section, we demonstrate that investment strategies with the potential to generate remarkably high yield also bear high risks. While our simulation only illustrate return courses in a deterministic fashion, through various simulated scenarios one can easily extrapolate that
the ever-changing market conditions---including volatile price movements and trading activities---lead to return instability, and sometimes even losses.
We discuss risks further in
\ref{sec:risks}.

\section{Current Major yield aggregators in DeFi}
\label{sec:comparison}

\begin{figure}[ht]
\centering
    \includegraphics[width=\linewidth]{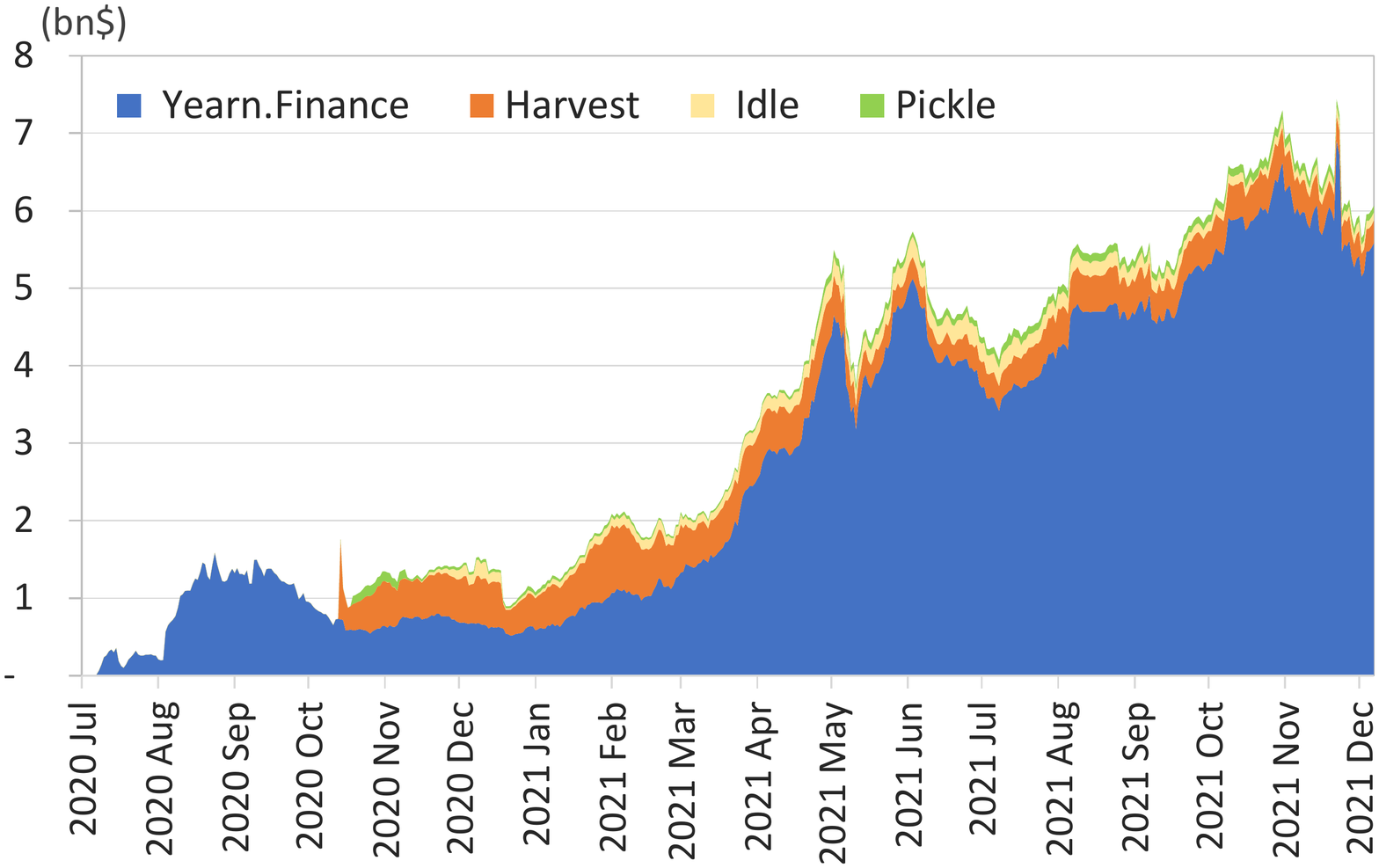}
\caption{Total value locked (bn\$). Data collected on 19 December 2021 from \url{https://defillama.com/} - Assets.
\label{fig:tvl}
}
\end{figure}

To date, yield aggregators have collected billions of dollars worth of liquidity. \autoref{fig:tvl} charts the evolution of total value locked for current major protocols since July 2020. This number indicates the total amount of funds that users deposited in different yield aggregators on Ethereum. 
This section compares current major yield aggregator protocols with a focus on strategies, performance and fee mechanism.

\autoref{tab:overview} shows the characteristics of the current major protocols and a few more. The data was collected on 19 December 2021. 

\begin{table*}[htbp]
\setlength{\tabcolsep}{4pt}
\caption{Overview of major existing yield aggregators}
\begin{tabularx}{\linewidth}{@{}X@{\hspace{1pt}}llllllllrlrrr@{}}
\toprule

& & \multicolumn{3}{c}{\ref{item:phase0}} & \multicolumn{3}{c}{\ref{item:phase2}} & & & & & & \\
& & \multicolumn{3}{c}{Supported pool assets} & \multicolumn{3}{c}{Strategies} & & & & & & \\
            \cmidrule(lr){3-5} \cmidrule(lr){6-8} 

\textbf{Protocol} &
& \textbf{Single} & \textbf{LP} & \textbf{Multiple} & \textbf{Simple} & \textbf{Leveraged} & \textbf{Liquidity} & \textbf{Pool} & \textbf{\# of} & \textbf{Governance} & \textbf{MCap*} & \textbf{TVL} & \textbf{Token}\\

& & \textbf{asset} & \textbf{token} & \textbf{assets} & \textbf{lending} & \textbf{borrow} & \textbf{provision} & \textbf{name} & \textbf{pools} & \textbf{token} & \textbf{(mm\$)} & \textbf{(mm\$)} & \textbf{holders}\\

\midrule

Idle Finance & \cite{idle2021} & \CIRCLE & \Circle & \Circle & \CIRCLE & \Circle & \Circle & Pool & 10 & \tokenaddress{IDLE}{0x875773784Af8135eA0ef43b5a374AaD105c5D39e} & 29.8 & 105.8 & 3,704\\
Pickle Finance & \cite{pickle2021} & \Circle & \CIRCLE & \Circle & \Circle & \Circle & \CIRCLE & pJar & 19 & \tokenaddress{PICKLE}{0x429881672B9AE42b8EbA0E26cD9C73711b891Ca5} & 9.2 & 88.5 & 7,238\\
Harvest Finance & \cite{harvest2021} & \CIRCLE & \CIRCLE & \Circle & \CIRCLE & \Circle & \CIRCLE & Vault & 53 & \tokenaddress{FARM}{0xa0246c9032bC3A600820415aE600c6388619A14D} & 64.3 & 295.5 & 12,003\\
Yearn Vaults & \cite{yearn2021} & \CIRCLE & \CIRCLE & \Circle &  \CIRCLE & \CIRCLE & \CIRCLE & Vault & 50 & \tokenaddress{YFI}{0x0bc529c00C6401aEF6D220BE8C6Ea1667F6Ad93e} & 1,156.6 & 5,548.1 & 44,484\\
Vesper & \cite{vesper2021} & \CIRCLE & \Circle & \Circle & \CIRCLE & \Circle & \Circle & Grow Pool & 4 & \tokenaddress{VSP}{0x1b40183efb4dd766f11bda7a7c3ad8982e998421} & 286.4 & 927.3 & 9,458\\
Rari Capital Earn & \cite{rari2021}  & \CIRCLE & \Circle & \CIRCLE & \CIRCLE & \Circle & \Circle &  Pool & 3 & \tokenaddress{RGT}{0xD291E7a03283640FDc51b121aC401383A46cC623} & 157.5 & 196.9 & 9,559\\
Value DeFi & \cite{valuedefi2021} & \CIRCLE & \CIRCLE & \CIRCLE & \CIRCLE & \Circle & \CIRCLE & vSafe & 4 & \tokenaddress{VALUE}{0x49E833337ECe7aFE375e44F4E3e8481029218E5c} & 36.2 & 959.5 & 4,928 \\
\bottomrule
\end{tabularx}
\scriptsize \vskip 1mm
Data fetched in December 2021.

In \ref{item:phase0}, some farming pools only accept single assets, 
some accept LP tokens,
and others accept more than one ERC20 token.

In \ref{item:phase2}, some protocols focus on lending strategies, others on leveraged borrowing or liquidity provision strategies. Some protocols employ multiple strategy groups.

* Fully diluted market cap: theoretical market cap which is calculated by assuming all the tokens were already in circulating supply.
\label{tab:overview}
\vspace{-6mm}
\end{table*}

\subsection{Idle Finance}

Launched in August 2019, Idle is a yield aggregator that automatically allocates and aggregates interest-bearing tokens \cite{idle2021documentation}. 



\subsubsection{Strategies}

Idle Finance only distributes single-asset pools over different lending protocols.
In \ref{item:phase0}, users' funds are pooled together and depending on the strategy that the pool employs, assets are allocated over different lending platforms, currently\footnote{\label{noteToday}\today} limited to: Compound, Fulcrum, Aave, DyDx and Maker DSR. There is no need for \ref{item:phase1} in Idle Finance, as the currently supported pools can be deposited directly into the above lending platforms. When any user interacts with Idle or if no interactions are made for 1 hour, rebalancing of the assets takes place according to the rates of supported providers.

Currently, Idle uses two different allocation strategies:
\begin{itemize}
    \item Best-Yield: this strategy seeks the best interest rates across multiple lending protocols.
    \item Risk-Adjusted: this strategy automatically changes the asset pool allocation in order to find an allocation with the highest risk-return score, compared to the highest return score of the \enquote{Best Yield} strategy. It does this by incorporating a framework for quantifying risk, developed by DeFiScore \cite{defiscore2021}, which outputs a 0-10 score that represents the level of risk on a specific lending platform (0 = highest risk, 10 = lowest risk).
\end{itemize}

\subsubsection{Return for users}

Idle uses {\tt IdleTokens} to represent the farmers' proportional ownership of the asset pool, which should accrue yield over time. In addition, farmers are rewarded with \coin{IDLE} governance tokens for participating in the pools as part of Idle's liquidity mining program. In January 2021, a two-year liquidity mining program started to reward liquidity providers depending on the amount of funds deposited and the utility generated by a certain pool \cite{idle2021distribution}.

\subsubsection{Protocol fees}

A performance fee 10\% of the generated yield is charged.


\subsection{Pickle Finance}

Launched in September 2020, Pickle offers yield on deposits through two products: Pickle Jars (pJar) and Pickle Farms. Jars are yield farming robots, earning returns on users' funds, while farms are liquidity mining pools where users can earn {\tt PICKLE} governance tokens by staking different kinds of assets. Proportional shares in pJars are represented by \coin{pTokens}.



\subsubsection{Strategies}

Each Pickle Jar employs a specific strategy to earn yield. Currently, two main versions of pJar strategy are in existence, pJar 0.00 and pJar 0.99, of which the 0.99 version is most important. There is no \ref{item:phase1} involvement in either version, pooled funds are directly utilized to farm rewards in \ref{item:phase2}, after which they are sold in \ref{item:phase3} to re-invest the accrued yield.

\begin{itemize}
    \item pJar 0.00: These pJars involve a user depositing LP tokens acquired by supplying liquidity on Curve Finance \cite{Curve2021}, an AMM-based DEX (see \ref{sec:amm}). The strategy employed in pJar 0.00 earns and re-invests \coin{CRV} rewards by selling \coin{CRV} into the market for stablecoins and re-depositing those into the Curve pools to get more LP tokens. Effectively, pJars 0.00 generate yield by accruing
    \begin{enumerate*}[label={(\roman*)}]
    \item LP fees from Curve and
    \item generating \coin{CRV} tokens because of Curve's liquidity mining program \cite{curve2021tokenomics}.
    \end{enumerate*}
    \item pJar 0.99: These pJars utilize LP tokens from Uniswap and Sushiswap, earning yield by accruing
    \begin{enumerate*}[label={(\roman*)}]
    \item LP fees from Uniswap/ Sushiswap and 
    \item generating \coin{SUSHI} or other native tokens because of liquidity mining programs.
    \end{enumerate*}
\end{itemize}

\subsubsection{Return for users}

Return of Pickle users is generally composed of
    \begin{enumerate*}[label={(\roman*)}]
    \item the fees accrued by providing liquidity to AMM pools,
    \item earning tokens distributed through external liquidity mining programs, and
    \item extra \coin{PICKLE} tokens if the yield farmer makes use of the Farm products. 
    \end{enumerate*}
    The return is thus dependent on underlying market forces, liquidity programs and token values. For example, the Pickle emission schedule defines how much \coin{PICKLE} will be distributed over time \cite{pickle2021emission}. 


\subsubsection{Protocol fees}

Most Pickle Jars have a 20\% performance fee on the generated yield.


\subsection{Harvest Finance}

Harvest Finance gives \coin{FARM} holders the opportunity to share in the revenue model of the protocol. By staking \coin{FARM}, users are entitled to receive part of the revenue that is collected by the protocol. Harvest Finance went live in August 2020, and currently has more than 
70 pools/vaults in its offering. 



\subsubsection{Strategies}

Harvest Finance has two main categories of yield farming strategies~\cite{harvest2021strategies}:

\begin{itemize}
    \item Simple single-asset Strategies: Users deposit single assets such as \coin{USDC}, \coin{USDT}, \coin{DAI}, \coin{WBTC}, \coin{renBTC} or \coin{WETH} into a Harvest Vault, which then deposits those assets into another yield generating protocol, including Compound and Idle Finance. 
    \item LP token Strategies: Users deposit LP tokens from Uniswap, Sushiswap or Curve into Harvest which automatically collects liquidity mining rewards and re-invests them into LP tokens.
\end{itemize}

\subsubsection{Return for users}

Depending on the vault used, return of Harvest users is composed of
\begin{enumerate*}[label={(\roman*)}]
\item the fees accrued by providing liquidity to AMM pools or other yield-bearing assets, 
\item earning tokens distributed through external liquidity mining programs and 
\item extra \coin{FARM} tokens as part of the liquidity mining program. 
\end{enumerate*}
These returns are dependent on underlying market forces, liquidity programs and token values. For example, the Harvest emission schedule defines how much \coin{FARM} will be distributed over time \cite{harvest2021token}.


\subsubsection{Protocol fees}

Harvest Finance does not charge withdrawal fees and does not claim a direct ``fee'' on the yield farming revenue. However, during liquidation of the yield, 30\% of the profits is used to buy the \coin{FARM} token on the market, which is then distributed to users who stake \coin{FARM} in the profit-sharing \coin{FARM} pool \cite{harvest2021vaultFunction}.\footnote{This type of buyback reduces the supply of governance tokens in the secondary market to the benefit of existing tokenholders.}


\subsection{Yearn Finance}
\label{subsec:yearn}

Yearn Finance offers a multitude of products in Decentralized Finance (DeFi), providing lending aggregation, yield generation and others \cite{yearn2021intro}. The services discussed here are Yearn Earn, a lending aggregator, and Yearn Vaults, a more comprehensive yield aggregator. Yearn Finance launched in July 2020.



\subsubsection{Strategies}

\begin{itemize}
    \item Earn pools: The strategy of the Earn pools is to collect a certain asset and deposit it either in dYdX, Aave or Compound, depending on where the highest interest rate of that asset is found. Yearn will withdraw from one protocol and deposit to another automatically as interest rates change between protocols. Earn pools thus omit \ref{item:phase1} and go directly to \ref{item:phase2}, in a strategy that is slightly similar to the Idle Finance ``Best-Yield'' Strategy. Proportional shares in Earn pools are commonly represented by \coin{yTokens}.
    \item Vaults: A Yearn Vault uses an asset as liquidity, deposits that liquidity as collateral (accounting for risk levels) to borrow stablecoins. Then, it uses those stablecoins to generate yield, after which that yield is re-invested in the stablecoins to generate more yield. Vaults thus employ all phases as illustrated in \autoref{fig:mechanism_all}, and allows for more complex strategies compared to Earn pools. Proportional shares in Yearn Vaults are commonly represented by \coin{yvTokens} or other \coin{yTokens}.
    
    
\end{itemize}

Yield farming strategies (\ref{item:phase2}) in Yearn v2 Vaults can be complex, involving flash loans (uncollateralized loans that are taken and repaid within the same transaction \cite{Xu2021c}), leveraged borrowing, staking on specific protocols (for example HegicStaking) and more.

\subsubsection{Return for users}

Yearn Finance distributed the \coin{YFI} governance token over a 9-day period after launch. Liquidity providers in the Earn pools or Vaults are thus not incentivized by a Yearn liquidity mining program, so current yield only comes from the returns that the product strategies reap. Those returns can be straightforward, as is the case for Earn pools, and can be complex to calculate, as is the case for v2 Vaults that can have up to 20 strategies working at once. Some Yearn vaults accept LP tokens, other accept single asset tokens.

\subsubsection{Protocol fees}

v1 Vaults have a 20\% performance fee and a 0.5\% withdrawal fee (in case funds need to be pulled from the strategy in order to cover the withdrawal request). v2 Vaults have also a 20\% performance fee, but no withdrawal fee. Instead, they charge a 2\% management fee. Performance fees are split 50:50 between the Treasury and the Strategist, the official creator of the strategy. The management fee is assigned fully to the Treasury.


\subsection{Empirical performance}

By fetching on-chain data of the above mentioned four main protocols, we visualized the evolution of selected vaults for comparison purposes. The evolution of a) total supply, and b) price per share are plotted for the \coin{DAI} vault, \coin{USDC} vault, and \coin{3crv} vault for aggregators where these specific pools are available. \coin{3crv} is the LP token name for the \href{https://curve.fi/3pool}{Curve Finance 3pool}. The name ``vaultTokens'' is used to represent the share of the corresponding Token-pool in a specific protocol. The price per share represents the ratio of underlying tokens and vaultTokens, which should be 1-to-1 pegged to a certain currency in the beginning.

\begin{figure*}[!tbp]
  \centering
       \begin{subfigure}{\linewidth}
        \centering
        \includegraphics[height=0.15\textheight]{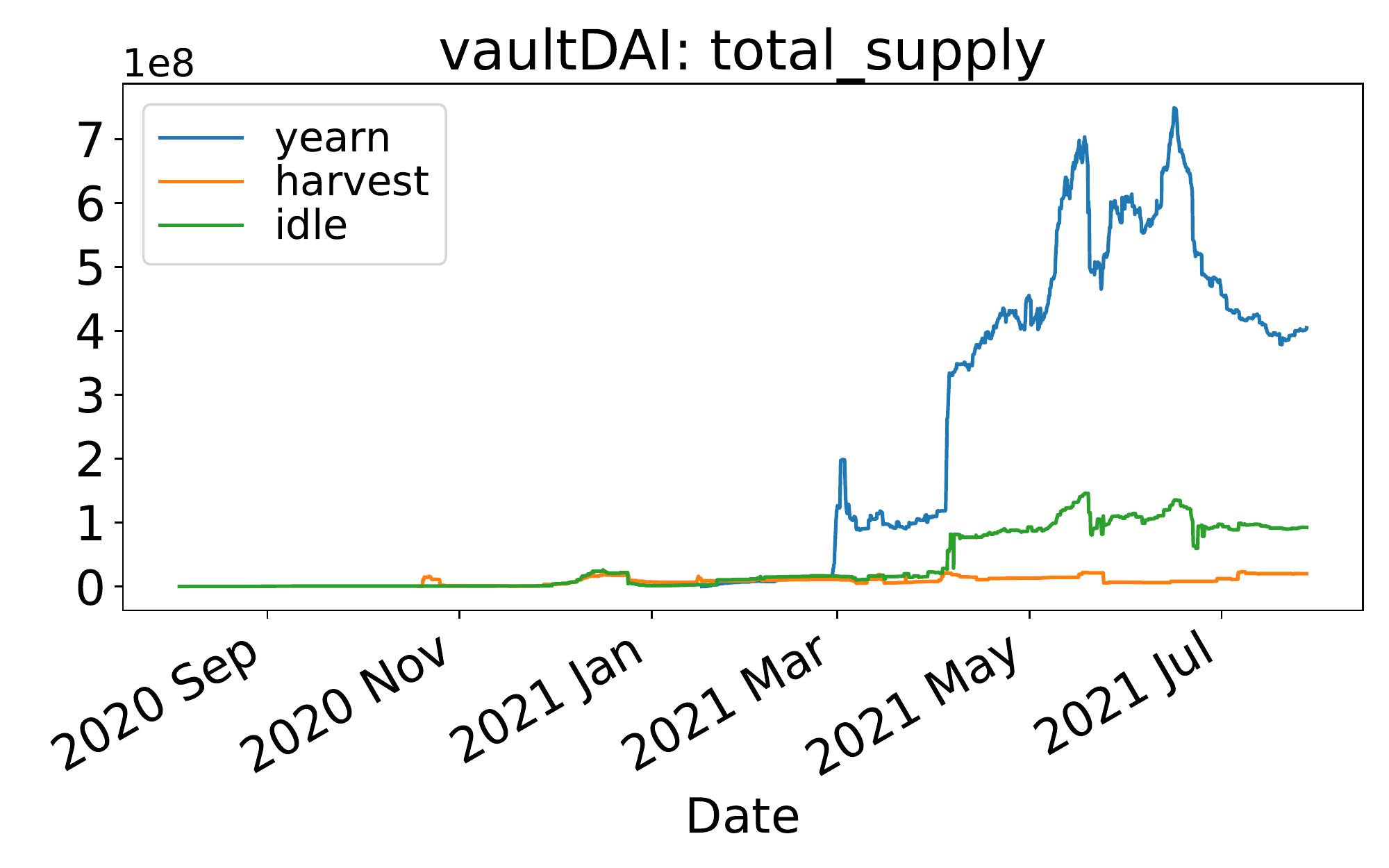}
        \includegraphics[height=0.15\textheight]{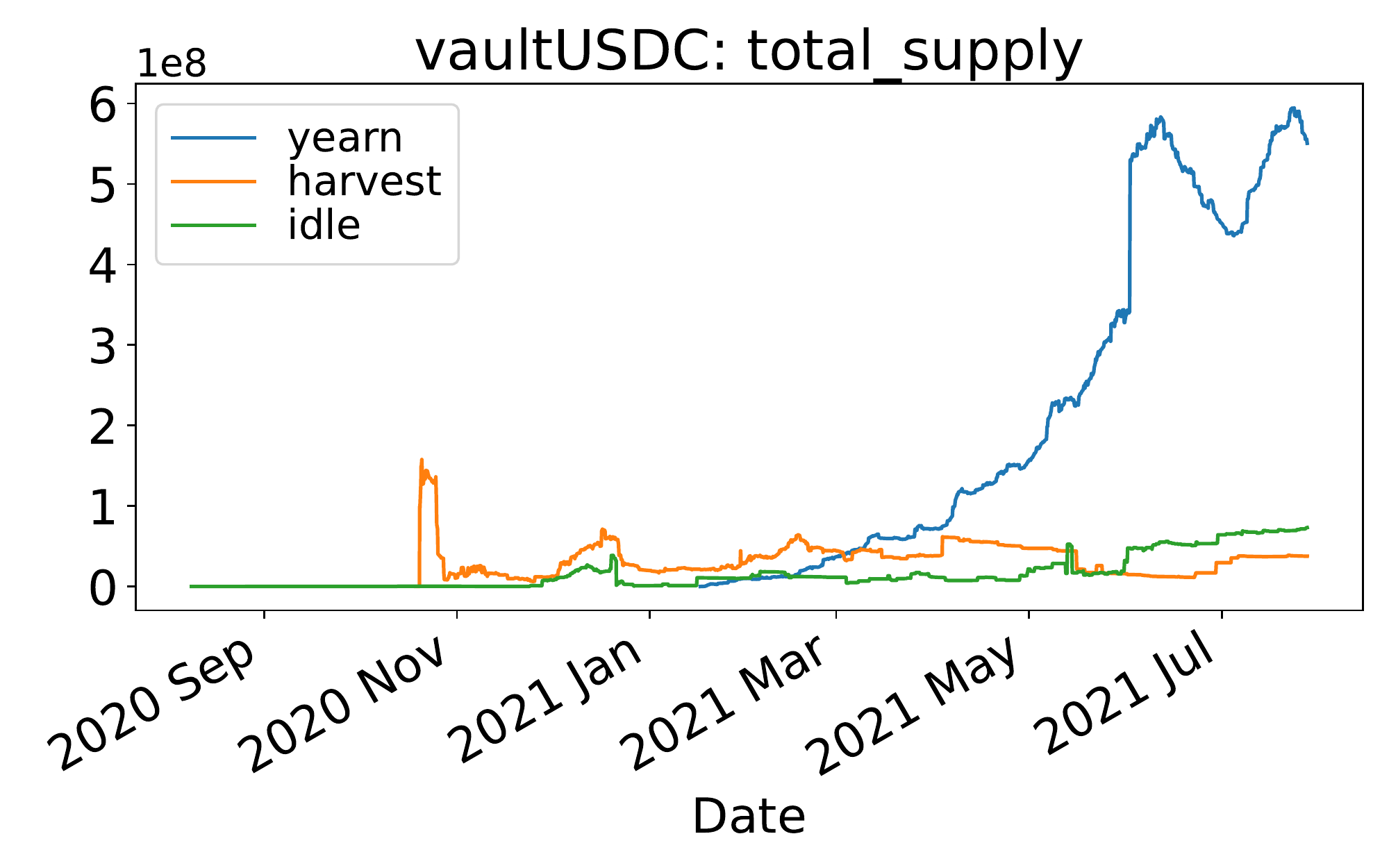}
        \includegraphics[height=0.15\textheight]{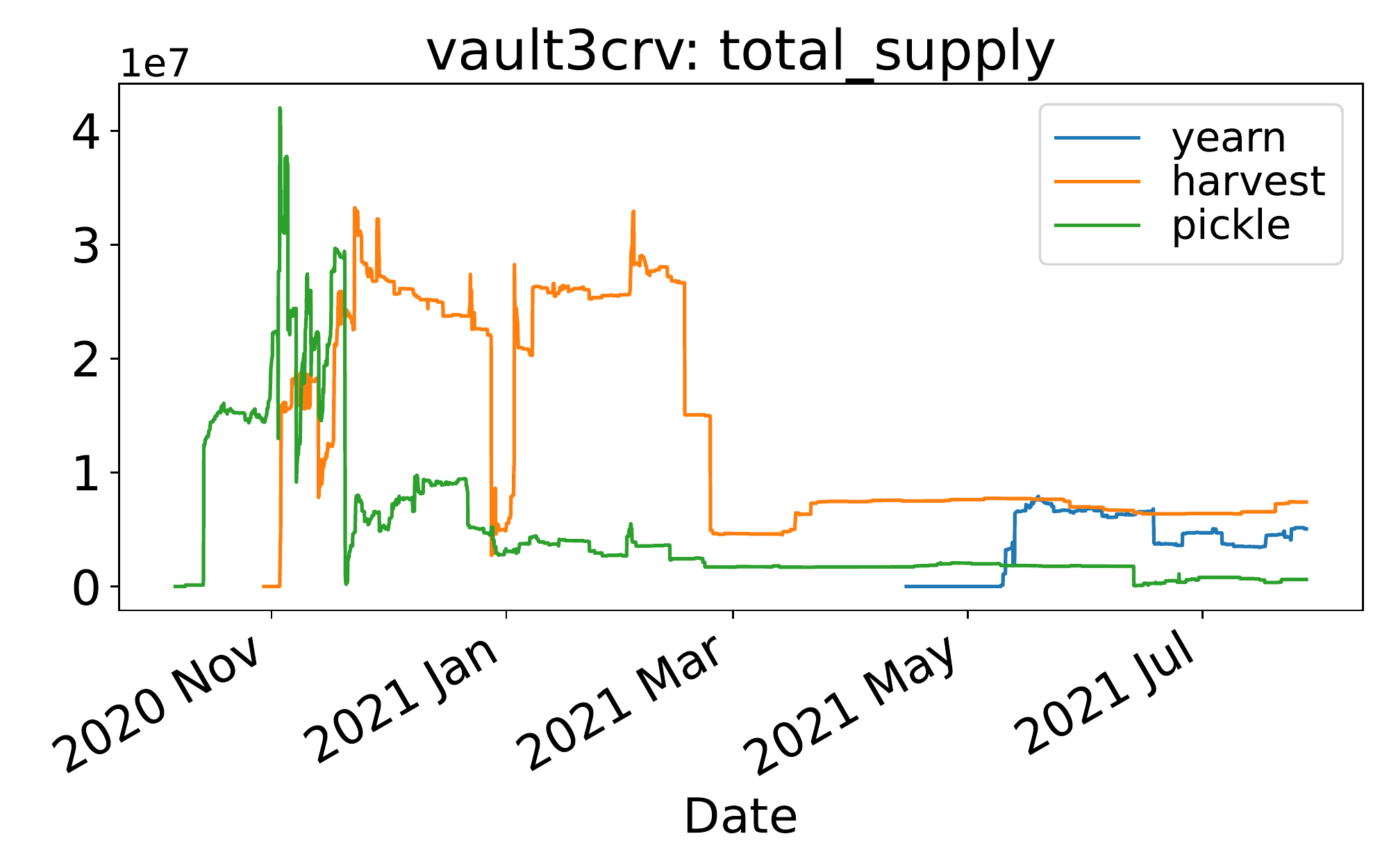}
        \caption{Total pool shares}
        \label{fig:total_supply_vault}
        \end{subfigure}
        
        \begin{subfigure}{\linewidth}
        \centering
        \includegraphics[height=0.15\textheight]{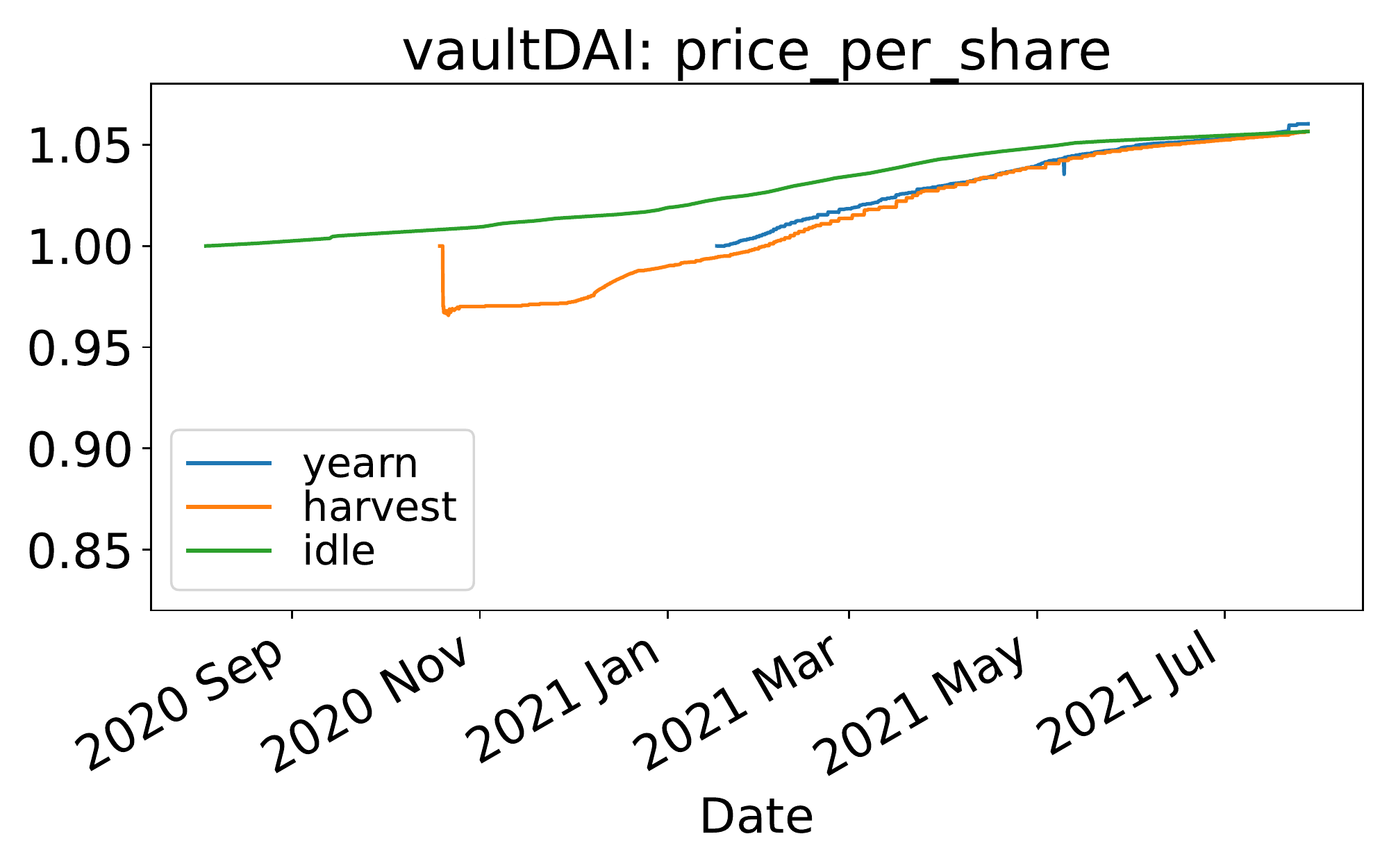}
        \includegraphics[height=0.15\textheight]{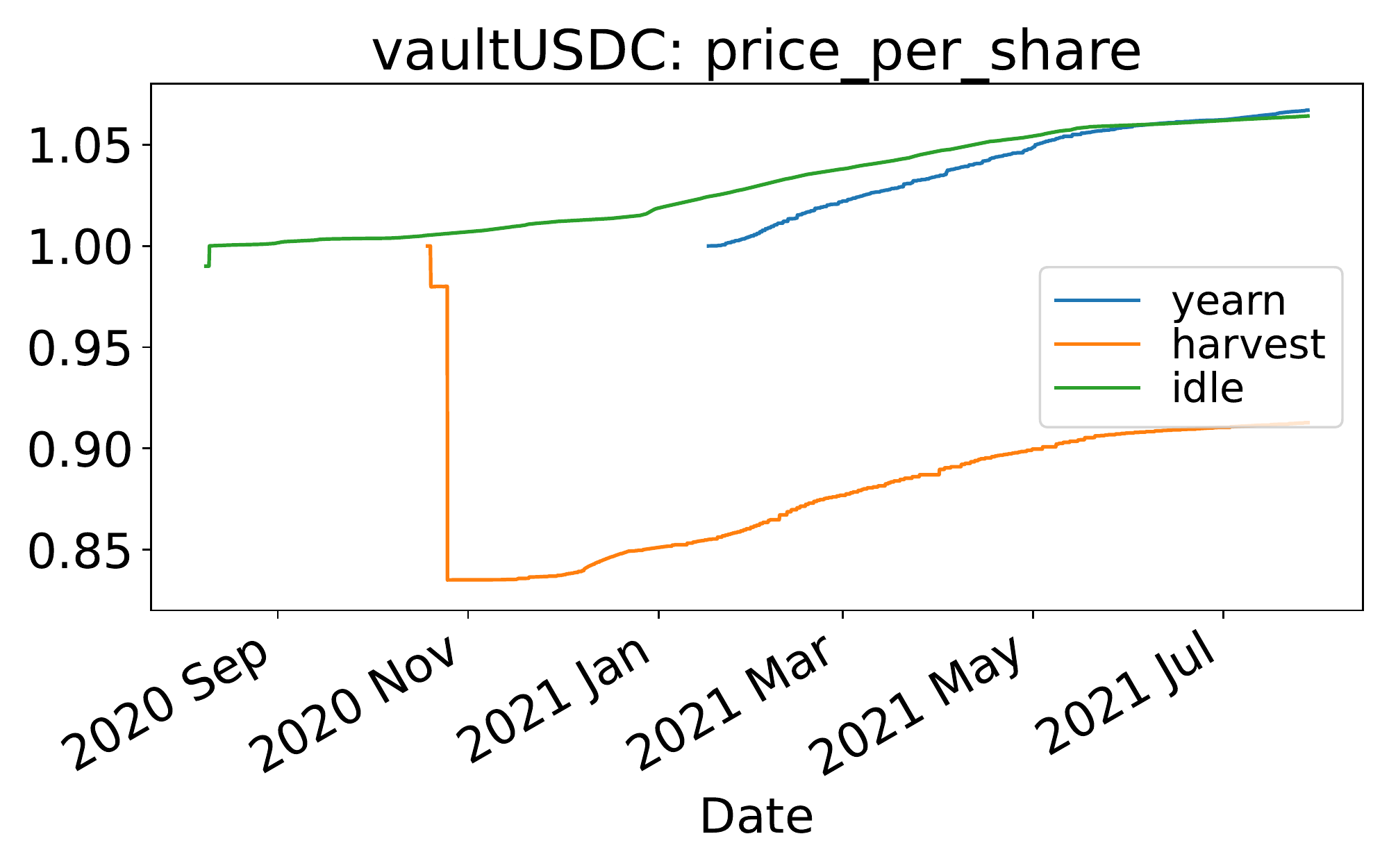}    \includegraphics[height=0.15\textheight]{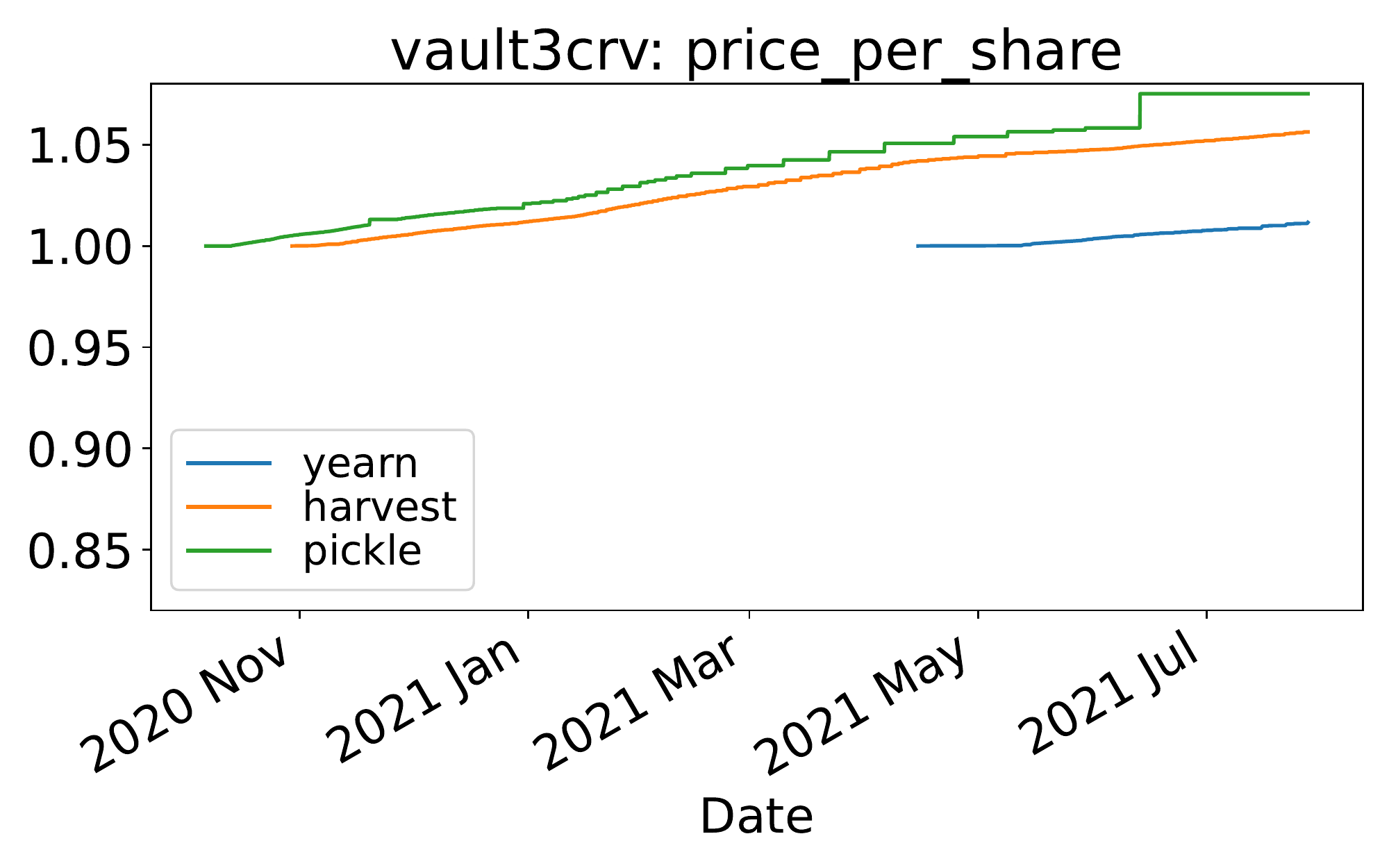}
        \caption{Price per share}
        \label{fig:price_per_share}
        \end{subfigure}

        \caption{Empirical data}
        \label{fig:empirical_results}
\end{figure*} Generally, the value of the vaultTokens then increases relatively compared to the underlying token as yield is accrued and added to the pool. Results are obtained by collecting the block where the vault is created, and calling the respective view functions with a block interval of 500 blocks until August 2021. The obtained timelines are shown in \autoref{fig:empirical_results}.

Regarding the total pool supply, Yearn shows an overall increase till June 2021 for all vaults, while those of the other protocols are either stable or volatile (especially in vault3crv).
Meanwhile, the surge in total supply for Harvest vaultUSDC in the middle of October 2020, followed by the drop at the end of the same month coincides with the surge in TVL of Harvest by 21st October (TVL achieved \$704.1 million on 21st October, up 110\% from \$334.41 million a week earlier), and the Harvest flash loan attack on 26th October, respectively.
In the attack, an attacker stole funds from the \coin{USDC} and \coin{USDT} vaults of Harvest Finance. A malicious actor had exploited an arbitrage and temporary loss, allowing this actor to obtain vault shares for a beneficial price, after which they were burned for profit \cite{HarvestFinance2020}. Looking at the vault3crv, drops in total supply of Harvest Finance are observed at the beginning of January 2021 and towards the end of February 2021 as well.

Concerning price per share, the number steadily increases overall, aside from the sharp declines observed for Harvest vaultDAI and vaultUSDC at the end of October 2020. These are again attributed to the flash loan attack that occurred on 26th October, where funds were drained out of the pool without equally affecting the number of vaultDAI. Further, the graph shows that the Yearn vaults update their contract relatively often, resulting in a short lifespan of plots---otherwise, a sharp decrease in price per share might have been observed after the attack against Yearn Finance on 4 February 2021, in which an unknown entity stole \$2.8 million, resulting in Yearn's \coin{DAI} vault losing \$11.1 million \cite{yearn2021exploit}.
The hacker(s) exploited the vault using Aave---a complex exploit with over 160 nested transactions and 8.6 million gas used (around 75\% of the block) reportedly resulted in a loss of \$2.7 million \cite{Etherscan}.


\subsection{Other aggregators}

The four main yield aggregators above are deemed the most mature, but a new wave of yield aggregating protocols is coming up. In general, there seems to be a tendency where more recent yield aggregators aim to be a one-stop-shop, providing additional functionalities such decentralized exchanges, lending and borrowing and risk-managing services. This enhances user experience and introduces more revenue streams for the protocols.

Below we list more recently launched protocols and products that are still being tested.

\paragraph{Vesper Finance}
Vesper \cite{vesper2021} 
focuses on institutional adoption of the DeFi yield market. Currently, only Vesper Grow Pools are available, which are comparable to the traditional yield products. In future developments, Vesper plans to integrate 
Vesper Labs \cite{vesper2021} where external users can build their own strategy in return for part of the reaped profits. 

\paragraph{Rari Capital}
Rari Capital \cite{rari2021} is a roboadvisor that attempts to provide investors with the highest yield, beyond just lending. It has multiple products, including Earn, Tranches, Fuse and Tanks. The Earn product can be considered a traditional yield farming service, while the other products are extending the number of functionalities on the Rari Capital platform, such as lending and borrowing, and yield farming within certain risk boundaries, called ``tranches'' \cite{rariTranches2020}.

\paragraph{Value DeFi}
Value DeFi \cite{valuedefi2021} has developed three main products: Value Liquid, Farming services and Vaults. Value liquid is an AMM platform, with vSwap and vPegSwap, allowing users to trade assets and earn fees if they provide liquidity in pools. The farming services, called vFarms, offer users staking possibilities with different returns. The vaults, called vSafes, are the more standard yield-generating contracts as discussed in this paper.

\paragraph{Alpha Homora}
Alpha Homora provides leveraged yield farming and leveraged liquidity provision. Users can participate in various roles in Alpha Homora, including Yield Farmers, Liquidity Providers, \coin{ETH} Lenders, Liquidators, and Bounty Hunters \cite{alpha2021docs}. This protocol does not specifically focus on extending the number of functionalities in one application, but more on allowing users to leverage their positions.

\paragraph{Bank of Chain}
Bank of Chain (BoC) offers automated yield farming services for stablecoins on EVM-compatible blockchains including Ethereum, Polygon and BNB Chain \cite{BankOfChain2022BankChain}. 
The protocol  
supports widely-adopted USD-pegged stablecoins such as \coin{DAI}, \coin{USDC} and \coin{USDT}. The BoC vault collects stablecoin deposits which are consequently invested in a range pre-selected yield-generating pools from other vetted protocols. Fund deployment, yield farming, reinvestment of yield, portfolio re-balancing are all automatically conducted following a pre-programmed strategy.

\subsection{Summary}

Many yield farming strategies entail some extent of optimization, e.g. choosing the lending pool that offers the highest APY to deposit assets into (e.g. Idle's Best-Yield pools, Yearn's earn pools), or balancing between risks and return (e.g. Idle's risk-adjusted pools). The core strategies applied by major yield aggregators commonly do not deviate much from the basic strategies described in \autoref{sec:formalization}. However, as the competition in yield farming grows, basic strategies becomes less effective \cite{Jakub}, which prompts protocols to device more sophisticated strategies that incorporate various forms of interactions with other DeFi protocols (e.g. upgrade from Yearn v1 to v2). Yield aggregators generate revenues by charging fees from investors. Protocols associated with better yield farming performance are able to charge higher fees, which can be observed by comparing both the performance fee between Yearn (20\%) and Idle (10\%) as well as their respective performance (see \autoref{fig:price_per_share}).


\section{Discussion}
\label{sec:discussion}

\subsection{Benefits of using yield aggregators}

The aggregation level offered by yield aggregators typically has a number of advantages. 

First, yield farmers do not have to actively compose their own strategy, but they can make use of the workflows invented by other users (called strategists), turning their investment strategy from active to a more passive approach without the need for extensive knowledge about the underlying protocols. Strategy creators, whether those are employed by the protocol or external stakeholders are constantly in search of new strategies that better captures the current DeFi opportunities. 

Second, because cross-protocol transactions are happening through a smart contract, capital shifts are done automatically, removing the need for the user to transfer funds manually between protocols. 

Finally, because funds are pooled in a strategy contract, the gas costs are socialized, resulting in fewer and thus lower interaction costs.
The disadvantage of aggregating multiple protocols together is the exposure to the multitude of risks when relying on a stack of products where each building block could be a potential liability in the process. Because these risks are an important consideration when deciding upon investing strategies, they are discussed separately in \ref{sec:risks}.

\subsection{Risks of using yield aggregators}
\label{sec:risks}

Traditionally, high rewards come with high risk and this is no different with yield farming. The composability of strategies allows creation of endless possibilities, yet in every element and layer of the stack, there are potential risks that should be accounted for. This section discusses the prominent risks in yield farming strategies. 

\subsubsection{Lending and borrowing risks}
Yield farming strategies might use lending and borrowing transactions as part of creating yield. Both come with risks. 

\paragraph{Liquidity risk}

In lending platforms, the Utilization rate $U$ is generally defined as $TotalAmountBorrowed / TotalLiquidity$. A utility ratio of $U = 0$ means that there is no demand for the supplied funds and there are no borrowers. We speak of ``under-utilization'', which is bad for business, as there is low interest. In this situation, there is no real risk. When utility ratio is close to $U = 1$, nearly all funds supplied in a pool are being borrowed. While this results in high interest rate, it also creates a so-called ``liquidity risk'': if many lenders withdraw at the same time---in a similar way to a bank run, a certain amount of them will have to wait until some of the borrowers have paid back their outstanding loans. Both lenders and borrowers are incentivized to get funds back into the pool because of a high premium rate, but in an high utilization scenario, there might be insufficient funds available for withdrawal. The liquidity risk certainly has a spillover effect: yield farming pools applying the lending strategy (see \ref{sec:simplelending}) also bear this risk, and can experience difficulty letting their investors withdraw their funds when the lending platform that has absorbed the funds is low in liquidity.

\paragraph{Liquidation risk}
\label{subsubsec:liquidationrisk}

Liquidation happens when the value of the collateral falls below a pre-determined liquidation threshold \cite{Perez2020liquidations}. In that case, the deposited collateral is no longer deemed valuable enough to cover the amount of the loan that was taken. At liquidation, the borrower losses part or all of the collateral, which the lending protocol automatically places for sale in the market at a discount with the proceeds used for loan repayment \cite{Werner2021sokDefi}. 
As should be clear from the above, using assets with high volatility in their price relative to the loaned assets increases the chance of liquidations. Yield farming pools that borrow tokens (see \ref{sec:leveragedborrow}) bear the liquidation risk. Depending on the liquidation threshold \cite{Aave2021a}---the loan to value ratio above which the liquidation is automatically triggered---the the borrower can lose up to half of their deposited assets.

\subsubsection{Composability risks}
Even though the composability factor of DeFi is what makes yield farming possible in the first place by allowing for complex, interconnected financial protocols, it does bring along the danger of smart contract risk as more and more money legos are plugged into a strategy. Both technical and economic weaknesses give rise to attractive exploit opportunities for malicious hackers. Funds in every step along the way can be compromised, potentially leading to huge losses for yield seekers. 

Individual smart contracts have the potential of containing software bugs, which could lead to unwanted behavior when interacting with the application. Many protocols try to mitigate such kinds of vulnerabilities by undergoing security audits by an external party or by launching bug bounty programs for community members.

Even if contracts are secure on an individual level, the combination of multiple smart contracts may not. The attack surface of a set of interacting smart contracts might be greater than the sum of its parts. Also, the effect of failure in one of the core components can cause a ripple effect throughout the whole ecosystem. Using a yield aggregator, the risk not only lies in smart contract risks of that aggregator, but also in all of the underlying protocols in lower layers.

\paragraph{Individual smart contract risk}

The implementation of a smart contract application has the potential of having software bugs in the source code, which could lead to unwanted behavior when using the application. 
Many protocols try to mitigate such kinds of vulnerabilities by undergoing security audits by an external party or by launching bug bounty programs for community members. 
While both initiatives are in no way a certainty for flawless execution afterwards, they provide another pair of eyes to the soundness of the contract.
Prudent yield aggregators tend to choose audited protocols for farming activities.

Furthermore, creating an application often involves design trade-offs, leading to weaker and stronger application aspects. For example, while the usage of an admin key for upgrading contracts results in more flexibility in improving the product, it also brings along the danger of a central authority deciding upon many product elements. 

Users can nowadays purchase cover from insurance protocols such as Nexus Mutual \cite{Cousaert2022} and InsurAce \cite{InsurAce.ioProtocol2021} to protect themselves against smart contract risks.

\paragraph{Composing multiple smart contracts risk}

While two protocols may be secure in isolation, the combination of them may not. By composing multiple smart contracts together, the attack surface might be greater than the sum of its parts. Also, the effect of failure in one of the core components can cause a ripple effect throughout the whole ecosystem. Thus, when using aggregator protocols, the risk not only lies in smart contract risks of that aggregator, but also in all of the underlying protocols in lower layers.
On 26 October 2020, an attacker executed a theft of funds from two Harvest Finance vaults. The attacker
manipulated the price of individual assets inside the Y pool of Curve, an AMM-based DEX where the funds of Harvest's vaults were invested, was able to steal 3.2\% of the total value locked in Harvest, worth \$33.8 million at that time \cite{HarvestFinance2020}. 

\subsubsection{APY instability}

Market forces can lead to instabilities in the returns, making the advertised APYs unreliable.

\paragraph{Volatile lending returns} 

While in \ref{subsec:examples} we assume a stable APY for deposits, the returns offered by lending protocols can be volatile in reality. It is important to note that interacting with a lending protocol can affect the utilization ratio, the key variable of an interest rate model.
In the case of a yield aggregator, funds are aggregated and moved in huge volumes. Lending returns can thus be significantly diluted with the market impact made by the funds deposited through a yield aggregator.

\paragraph{Divergence loss}

A commonly known risk of providing liquidity in AMMs is divergence loss, also known as impermanent loss (IL), caused by price volatility of the assets that were used to obtain LP tokens. Xu et al \cite{xu2021dexAmm}, explain this concept more in depth.
When one compares the value of holding specific assets outside of an AMM pool with supplying those same assets to a liquidity pool, there might be a value decrease because of price movements and subsequent arbitrage opportunities.
This loss is \enquote{impermanent} because as the exchange rate between the two assets in an AMM pool moves up and down, the loss experienced by LP constantly recur and re-disappear. 
loss is only realized when assets are actually taken out of the pool \cite{xu2021dexAmm}. 

Strategies that make use of liquidity provision techniques (see \ref{sec:liquidityprovision}) should take into account the risk of value decreases in supplied liquidity, as it can greatly influence the returns. One way to avoid this is to make use of protocols with some degree of IL protection, such as Curve, which implicitly limits impermanent loss by pooling assets with equal pegs, or Balancer that allows for weighted pool assets such that liquidity providers can retain most of the upside in case of price spikes.

\paragraph{Low trading activity}

When a strategy supplies part of the funds in a liquidity pool of a DEX (see \ref{sec:liquidityprovision}), the yield relies on trading fees which are subject to the activity within that pool. Low activity in the pool results in lower trading fees, decreasing APY. 

\paragraph{Price fluctuations in liquidity incentives}

As discussed in \ref{subsec:native_tokens}, protocols might reward users with liquidity incentive tokens. Demonstrated in \ref{subsec:examples}, strategies making use of this reward system are subject to the value of those incentives. Price fluctuations in those tokens cause APYs to be unstable and thus unreliable.

\paragraph{Unstable transaction cost}

Yield farming involves interaction with complex smart contracts containing numerous computations and operations. This implies, on EVM-compatible blockchains where most DeFi protocols operate, high gas consumption for farming transactions. The gas price typically increases when the network becomes more congested, and decreases when there is less traffic. Therefore, the transaction cost of yield farming activities largely depends on the network condition, which can be fairly volatile, contributing to the instability of yield APY.

\subsection{The sustainability of yield}

In the past years, a multitude of yield aggregator protocols have sprung up and while the general framework behind them is similar, they all have their own flavor. Idle Finance started in 2019 with a first version of their product, which deposited funds into the lending platform that gives the best rate at a given time. Inspired by the liquidity mining program of Compound, Yearn Finance extended this model in July 2020 by inventing more complex strategies, called Vaults, next to their Earn product. As more forms of liquidity mining programs sprung up, Harvest Finance and Pickle finance specialized in yield farming with LP tokens later that summer. 

It is not surprising that this all happened within a relatively short period of time. After all, in a decentralized world, developers have the opportunity to work on an open-source code base and improve the existing protocols. It is clear that instead of trying to improve or add features to a working product, developers choose to take parts of a existing product and create their own flavour of it. This is why most yield aggregators share a common base of code and workflows are very similar across the board.

This raises the question if we will see convergence or further divergence in the future of these protocols. The empirical data shows that in the early days, protocols like Harvest Finance and Pickle Finance were able to decrease the market dominance of Yearn Finance, in terms of TVL. Recently, Yearn Finance has taken the crown, with over \$4bn locked in the protocol. Idle, Harvest and Pickle Finance have a combined TVL of about \$550M. Vesper Finance, as the new kid on the block, has been performing exceptionally well, with currently about \$1bn locked. While it can be argued that there is a converging trend in terms of fund concentration, Vesper shows that new players still have the potential to disrupt. 

Furthermore, there seems to be a converging trend in the offered functionalities of the major protocols. Yearn Finance has a full product suite, of which the Vaults are only one part. Vesper Finance, Rari Capital and Value DeFi are building one-stop-shop solutions, offering a myriad of strategies for a multitude of assets, in-house exchanging services and lending/borrowing opportunities. Even though this paper focuses on the Ethereum blockchain, a similar tendency can be seen on the Binance Smart Chain, where protocols like PancakeSwap \cite{pancakeswap2021} offer multiple products in one place.

Yield aggregators have been and still are an attractive way to collect yield in DeFi. But how sustainable is this yield? Discussed in \ref{sec:yield_origin}, yield comes from three main sources. While research on the sustainability of yield deserves a separate examination, it could be argued that yield coming from native token distribution is relatively short-lived. Once emission schedules are finished, this yield source is depleted. Even though new protocols can blossom with their newly started emission schedules, it seems unlikely that this source of yield is sustainable. The demand for borrowing could be more sustainable in that regard, but it is highly dependent on market sentiment, especially for non-stablecoins. Yield from revenue sharing tokens seems to be the most durable, especially if DeFi is able to hold on to recent growth rates.

In the end, investors choose a yield aggregator that offers the most advantages while simultaneously diminishing the risks attached to this. This generally requires a trade-off between the costs of using a strategy, the security risks and the offered yield.

\section{Related work}

The investigation of DeFi protocols is a fairly new field, there is a paucity of existing works related to DeFi, especially the ones associated with yield aggregating protocols.
We present the previous studies related to our paper by dividing them into the following categories.

\subsection{DeFi protocols}

\cite{moin2019stablecoin} and \cite{Pernice2019a} systematically study the general designs of DeFi platforms by decomposing the structure into diverse elements, i.e., peg assets, collateral amount, price and governance mechanism, they investigate the pros and cons of DeFi platforms to spot future directions.

On the other hand, recent existing works mainly focus on particular types of DeFi protocols or explore unusual behaviours observed. 
For instance, Liu et al. \cite{Liu2020a} conduct a comprehensive measurement study of DeFi oracles of four prime DeFi platforms to find the operational issues intrinsic in oracles and common divergence between the real and achieved prices.

There are various papers relating to flash loans. Wang et al. \cite{Wang2020b} reveal a structure that enables the identification and classification of speculative flash loan transactions. Within the scope of the analysis of financial attack vectors that involve a flash loan, \cite{Qin2020c} study the existing flash loan-based attacks and propose optimizations that significantly improve the ROI of these attacks. Further, \cite{Gudgeon2020c} shows with specific focus on the leading DeFi project by market share, that it is feasible for attackers to combine crowdfunding and flash loans to execute a successful attack.

The mechanics and properties of DeFi lending protocols are investigated in several studies \cite{Gudgeon2020PLF,Bartoletti2020sokLendingPools,Perez2020liquidations,Tolmach2021,Kao2020a}.
Gudgeon et al. \cite{Gudgeon2020PLF} coin the phrase Protocol for Loanable Funds (PLFs), to name protocols that form distributed ledger-based markets for loanable funds. Further, they study the structure used to settle interest rates on three major PLFs, and provide several empirical analysis on the interest rate movements and the market efficiency. 

Structured models for lending protocols and
the relevant pools are developed in the following papers \cite{Bartoletti2020sokLendingPools,Perez2020liquidations,Tolmach2021}.
Bartoletti et al. \cite{Bartoletti2020sokLendingPools} present the major properties of lending pools and their synergies with other DeFi protocols. 
An empirical analysis of liquidations on lending platforms is conducted by Perez et al. \cite{Perez2020liquidations}, by utilizing the abstract model of Compound. This work employs a more generic model of a pool, enabling both lending and DEX protocols to be formalized.
Tolmach et al. \cite{Tolmach2021} further set formal definitions for the dominant DeFi components and propose a formal process-algebraic technique to model DeFi protocols to enable efficient property verification.

With regards to studies that utilise simulation and stress tests, Kao et al. \cite{Kao2020a} employ agent-based modelling and simulation-based stress tests, in order to evaluate the economic security of the Compound protocol. Lewis et al.
\cite{Gudgeon2020c} explore how design weaknesses in DeFi protocols can trigger a decentralized financial crisis, by for instance conducting stress tests with Monte Carlo price simulation to show how a DeFi lending protocol may find itself undercollateralized. 


\subsection{Fund management}
The {\it portfolio optimization}, or portfolio management problem, which concerns the determination of the best portfolio (asset distribution) out of all sets of portfolios available under certain constraints, is regarded as the core problem of the field of fund management.

A fundamental approach to solve this question is posed by Markowitz \cite{1952Markowitz}, in which he advocates the mean–variance model to formulate the problem as an optimization problem entailing two criteria:
\begin{enumerate*}[label={(\roman*)}]
\item the reward (measured basically by return, and should be maximized),
and
\item the risk (measured by the variance of return, and should be minimized).
\end{enumerate*}
There does not exist a single optimal solution (portfolio), as there are two criteria to consider, but a set of optimal portfolios, i.e. {\it efficient portfolios}.

Putting the theory of Markowitz at the core, a myriad of papers have been issued by enhancing or reshaping the basic model in the following three ways \cite{2010Anagnostopoulos_portofolio}:
\begin{enumerate*}[label={(\roman*)}]
\item the simplification of the size and type of input data;
\item the replacement of the measures of risk;
and 
\item the introduction of further criteria or constraints.
\end{enumerate*}

\section{Conclusion}
\label{sec:conclusion}

In this SoK, we propose a general framework for yield farming strategies. First, we explain where yield comes from and describe a number of primitives in yield farming. We study protocols and tokens used by aggregators to generate yield, after which we combine this information to create a general workflow of yield farming strategies.
Second, we draft three examples of frequently used strategies and simulate yield farming performance under a set of assumptions.
Third, we compare four major yield aggregators by summarizing their strategies and revenue models and by evaluating on-chain data on three specific vaults.
Finally, we discuss the benefits and risks of yield aggregating protocols, together with related work in the DeFi industry.

While yield farming has been exploding since 2020, an important question remains if current yields will be sustainable in the long term. Higher rewards also imply higher risks, and DeFi attacks might prove that the safest and most robust yield provider might win the race. New industry developments focus on building one-stop-shop solutions, in pursuit of aggregating more than just yield and facilitating the on-boarding of new DeFi users. 

\newpage


\bibliographystyle{IEEETranS}
\bibliography{references}

\end{document}